\documentclass[preprint,superscriptaddress]{revtex4-1}

\usepackage[english]{babel}
\usepackage[ansinew]{inputenc}
\usepackage{boldline,multirow,xcolor,colortbl}
\usepackage{times}
\usepackage{graphicx}
\usepackage{graphics}
\usepackage{amsmath}
\usepackage{amsfonts}
\usepackage{amssymb}
\usepackage{amsbsy}
\usepackage{dsfont}
\usepackage{epstopdf}
\usepackage{makeidx}
\usepackage{subfigure}
\usepackage{color} 
\usepackage{pgf}
\usepackage{bm}
\usepackage{tikz} 
\usepackage[scr]{rsfso}
\usepackage[normalem]{ulem}
\usepackage{fancyhdr}
\usepackage{float}

\setcitestyle{super}

\begin{document}

 \newpage

\title{Quantum Frequency Combs with Path Identity for Quantum Remote Sensing} 

\author{D.A.R. Dalvit$^*$}
\affiliation {Theoretical Division, Los Alamos National Laboratory, Los Alamos, NM 87545, USA}

\author{T.J. Volkoff}
\affiliation {Theoretical Division, Los Alamos National Laboratory, Los Alamos, NM 87545, USA}

\author{Y.-S. Choi}
\affiliation {Center for Integrated Nanotechnologies, Los Alamos National Laboratory, Los Alamos, NM 87545, USA}

\author{A.K. Azad}
\affiliation {Center for Integrated Nanotechnologies, Los Alamos National Laboratory, Los Alamos, NM 87545, USA}

\author{H.-T. Chen}
\affiliation {Center for Integrated Nanotechnologies, Los Alamos National Laboratory, Los Alamos, NM 87545, USA}

\author{P.W. Milonni}
\affiliation {Department of Physics and Astronomy, University of Rochester, Rochester, NY 14627, USA}

\maketitle

%%%%

%%%%
\newpage

{\bf Quantum sensing promises to revolutionize sensing applications by employing quantum states of light or matter as sensing probes. Photons are the clear choice as quantum probes for remote sensing because they can travel to and interact with a distant target. Existing schemes are mainly based on the quantum illumination framework, which requires a quantum memory to store a single photon of an initially entangled pair until its twin reflects off a target and returns for final correlation measurements. Existing demonstrations are limited to tabletop experiments, and expanding the sensing range faces various roadblocks, including  long-time quantum storage and
photon loss and noise when transmitting quantum signals over long distances. We propose a novel quantum sensing framework  that addresses these challenges using quantum frequency combs with path identity for remote sensing of signatures (``qCOMBPASS"). The combination of
one key quantum phenomenon and two quantum resources,
namely quantum induced coherence by path identity, quantum frequency combs,
and two-mode squeezed light, 
allows for quantum remote sensing without requiring a quantum memory.
The proposed scheme is akin to a quantum radar based on entangled frequency comb pairs that uses path identity to detect/range/sense a remote target of interest by measuring pulses of one comb in the pair that never flew to target, but that contains target information ``teleported" by quantum-induced coherence from the other comb in the pair that did fly to target but is not detected.
We develop the basic qCOMBPASS theory, 
analyze the properties of the qCOMBPASS transceiver, and introduce the qCOMBPASS equation -- a quantum analog of the well-known LIDAR equation in classical remote sensing. We also describe an experimental scheme to demonstrate the concept using two-mode squeezed quantum combs. 
qCOMBPASS  can strongly impact various applications in remote quantum sensing, imaging, metrology,
and communications. 
These include detection and ranging of low-reflectivity objects, measurement of small displacements of a remote target with precision beyond the standard quantum limit (SQL), standoff hyperspectral quantum imaging, discreet surveillance from space with low detection probability (LPD, detect without being detected), very-long-baseline interferometry, 
quantum Doppler sensing,
quantum clock synchronization, and networks of distributed quantum sensors.
}
%%%%%%%%%%
\vspace{0.2cm}
\section{Introduction}

Classical remote sensing has a long history whose most notable representatives are RADAR and LIDAR technologies. Sensing with quantum light can overcome classical limitations in resolution and sensitivity in microscopy, spectroscopy, and imaging, and offers clear advantages over atomic sensors in applications such as remote sensing that require the probe to travel and interact with a distant target. Furthermore, sensing with low-flux quantum light is beneficial to analyze fragile biological specimens 
that would otherwise be destroyed by normal laser powers. It is also conducive to detect low-reflectivity objects in bright electromagnetic backgrounds. The standard method in photonic quantum sensing
is to create photon pairs in nonlinear crystals via the process of spontaneous parametric down-conversion (SPDC). One photon in the pair (the ``signal" photon) is sent to a target and partially reflects off target, and the other photon (the ``idler'') is stored in a quantum memory. Finally, correlation measurements are performed between the reflected and stored photons. The quantum memory can be an atomic cloud where the photon is stored, an optical fiber, or simply an optical cavity. This scheme, known as quantum 
illumination (QI) \cite{Lloyd2008,Tan2008,Maccone2020}, was demonstrated in table-top experiments in the 
optical \cite{Lopaeva2013,Zhang2013,Gregory2020} 
and microwave \cite{Barzanjeh2015,Barzanjeh2020}
regimes, and showed a minimal quantum advantage $<4$ dB over detecting the presence of an object using a coherent state of light (standard laser). 
Another well-known scheme for quantum sensing  is quantum ghost imaging (QGI) \cite{Strekalov1995}
, that uses momentum correlations between the photon pairs to image an object by measuring photons that never interacted with the target. A third quantum sensing scheme is based on quantum-induced coherence by path identity (QICPI) \cite{ZWM1991,Hochrainer2022}, which is at the core of our work. 
This remarkable phenomenon of quantum optics was discovered in 1991 by Zou, Wang, and Mandel (ZWM)  \cite{ZWM1991}, and in recent years the ZWM effect has experienced a revival of interest due to advances in quantum imaging \cite{Lemos2014,Qian2023}, sensing \cite{Kalashnikov2016}, and high-dimensional entanglement generation \cite{Kysela2020}. 
QICPI also uses photons pairs and can sense/image an object by measuring photons that never interacted with the target. Although it is similar to QGI on the surface, QICPI and QGI are drastically different. Quantum ghost imaging has been shown to have a classical analog using incoherent thermal light
\cite{Gatti2004}, while quantum-induced coherence by path identity is based on single-photon interference and, as such, is classically forbidden. To date, demonstrations of quantum sensing with QI, QGI, or QICPI have been limited to table-top experiments and used quantum memories to store single photons, either in the form of delay lines or cavities. Typical memories can coherently store single photons for only up to a few milliseconds  \cite{Yang2015}, limiting the total distance that the other photon in the pair can travel before coherence is lost to a few tens of meters. This constitutes a substantial roadblock for quantum remote sensing.

The paper is organized as follows. In Section II we introduce the qCOMBPASS concept.
Section III describes qCOMBPASS in the perturbative SPDC regime of low-flux pump where each generated signal/idler pulse has at most one photon.
Section IV discusses the ``chopper protocol" needed in qCOMBPASS to attain quantum-induced coherence by path identity.
Section V
contains the description of non-perturbative qCOMBPASS that occurs for high-flux pump and for which each generated signal/idler pulse pair is in a two-mode squeezed vacuum state.
The implementation of path identity in qCOMBPASS is analyzed in Section VI.
Section VII discusses the qCOMBPASS transceiver state and the associated two-mode
squeezed vacuum Wigner distribution. 
Realistic limitations of qCOMBPASS are contained in the following sections.
Section VIII considers light diffraction through the atmosphere and free space, and derives the distance dependence of the reflected signal field. 
Section IX contains a discussion of the effect of atmospheric turbulence on our sensing protocol.
Section X contains the key result of this work, namely the  qCOMBPASS equation that governs the rate of photo-detection.
Effects of signal photon absorption, squeezing level, photon collection efficiency, and detector efficiency, are taken into account. 
Quantum metrology using qCOMBPASS is briefly described in Section XI.
Section XII contains a description of possible experimental implementations of
the qCOMBPASS concept. Finally, the conclusions of this work
are summarized in Section XIII.

%%%%%%%%%%
\vspace{0.2cm}
\section{The qCOMBPASS concept} 

\begin{figure}[t]
\includegraphics[width=1\linewidth]{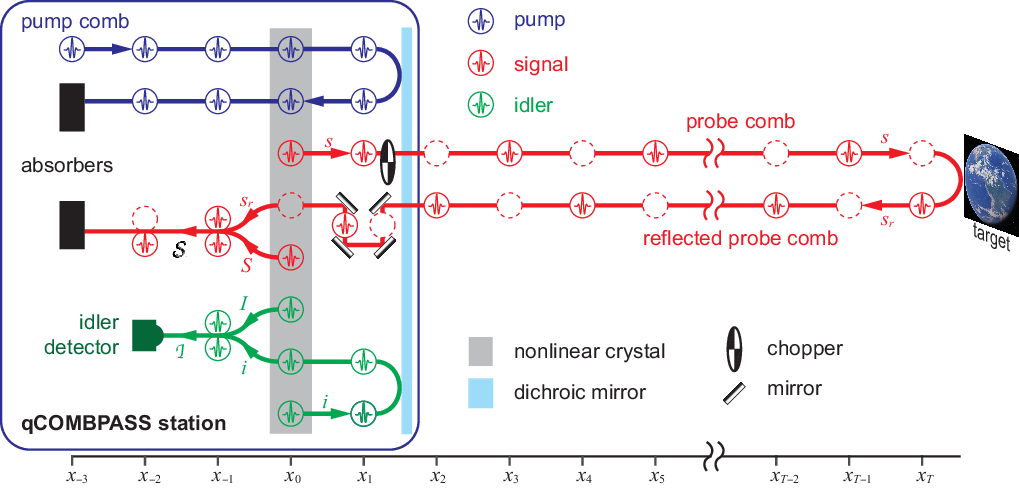}
\caption{{\bf The qCOMBPASS concept.}
Stroboscopic snapshot of qCOMBPASS spatiotemporal evolution.
Frequency comb pump makes a forward pass through a nonlinear crystal and
generates entangled pairs of quantum frequency combs with signal $s$  and idler $i$ pulses. Pump, signal, and idler combs are vertically displaced for clarity.
A chopper blocks every other signal pulse from going to a remote target (dashed circles). 
Dichroic mirror retro-reflects pump-idler combs and transmits
a signal comb.
In the reverse pass, the pump comb generates second entangled combs with signal $S$ and idler $I$ pulses. 
Two Y-junctions perform path identities $s_r, S \mapsto {\cal S}$ and
$i, I \mapsto {\cal I}$. No quantum memory is used to store any pulse.
}
\label{Fig1}
\end{figure}

The discovery of frequency combs - periodic trains of optical pulses with long intra-comb coherence times - provided a new tool for high precision classical remote sensing and imaging \cite{Trocha2018,Suh2018,Riemensberger2020}. Their recent generalization to the quantum domain offers new sources of quantum light for applications in quantum information science and technology
\cite{Reimer2016,Kues2017,Guo2017,Pasquazzi2018,Guidry2022}. 
qCOMBPASS uses special quantum combs as sensing elements and QICPI as the sensing protocol to enable quantum remote sensing. Two-mode squeezed vacuum  \cite{Schnabel2017} quantum frequency combs (TMSVc) and induced-coherence by path identity form the qCOMBPASS
transceiver state. The signal mode of a TMSVc is used as transmitter and the idler modes of this and a separate TMSVc function as a receiver. Pulses from the first comb propagate to a reflective target and acquire information, e.g., the position or material composition of the target, and return to the qCOMBPASS station after reflection from the target (Fig. \ref{Fig1}). In contrast to the quantum illumination  framework or other schemes based on reflected power, qCOMBPASS does not measure reflected photons nor does it perform any correlation measurement between reflected and stored photons. Instead, target information is retrieved from new idler pulses of two TMSVc that never flew out to the target and never physically interacted with reflected probe pulses. This is possible thanks to the phenomenon of quantum-induced coherence by path identity that imprints  information from the probe comb pulses to idler comb pulses. 
As will become clearer in the following, this is possible because qCOMBPASS creates an indistinguishability between an ``old" generated reflected photon and a ``new" locally generated photon.
Although both qCOMBPASS and QI transmit the signal mode of two-mode squeezed vacuum light to a reflective target, the state that is measured at the end of a QI protocol is not entangled. Specifically, in the operating regime in which QI exhibits a quantum advantage over a laser transmitter, entanglement between the transmitted mode and the quantum memory is destroyed \cite{Tan2008}. In sharp contrast, the quantum state measured in a qCOMBPASS protocol is an entangled state which is locally generated but contains information from the transmitted mode due to implementation of path identity. 

%%%%%%%%%%%%%%%
%%%%%%%%%%%%%%%

The ZWM phenomenon is rooted in single-photon interference, akin to the famous Young's double-slit experiment with single photons\cite{Taylor1909}. As a result, qCOMBPASS has no classical analog and is a classically forbidden sensing scheme, 
quantum supremacy over any classical remote sensing scheme based on classical light. In the regime where QI exhibits a maximal 6 dB quantum advantage over a laser transmitter, entanglement between the transmitted mode and the quantum memory (which is required for QI) is destroyed. The quantum state measured in a qCOMBPASS protocol, however, is an entangled state which is locally generated by spontaneous parametric downconversion (SPDC) and contains information from the transmitted mode due to the implementation of path identity. Furthermore, qCOMBPASS can potentially offer quantum advantage over QI and other classical sensing schemes in detection error in operating regimes of both low and high background noise (e.g., in optical and microwave domains) because  the quantum state of the qCOMBPASS transceiver is a path-identity-induced 
combination of Gaussian and non-Gaussian states generated from photon subtraction from the squeezed vacuum \cite{Fan2018, Ra2020,Walschaers2021}.
Finally, due to the amplification implemented at the qCOMBPASS receiver via the reverse quantum comb, one can anticipate order-of-magnitude reductions in the number of received photons necessary to achieve quantum advantage. Thus, even a simple photodetection scheme can allow advantage in detection error over a laser transmitter.

As already mentioned, quantum memories are currently limited, even with elaborate techniques, to sub-second storage times, posing a serious challenge to quantum remote sensing with QI. The key feature of qCOMBPASS, compared to QI and other techniques based on ZWM for sensing and imaging, is that it does not require a quantum memory to store pulses of the local quantum comb over a time it takes for their corresponding pulses of the transmitted comb to return from the target. This is due to the fact that a pulse in one comb is in phase with any other pulse in the same comb and with any other pulse belonging to the twin comb. Coherent pulses of a quantum frequency comb thus act in effect as a quantum memory. Hence, the path of a returning ``old" pulse belonging to one comb reflected from target can be identified with the path of a ``new" pulse of the locally generated twin comb, thereby resulting in quantum-coherent exchange of the information contained in the old pulse to the new pulse. Quantum interference occurs between pulses
 that were generated at substantially different times: one long in the past (the reflected pulse from target),
 one recently (a retro-reflected idler pulse), and two just generated (the signal/idler pulse pair generated by a pump pulse in reverse motion), see Fig.1.  The quasi-perfect intra- and inter-coherence properties of quantum frequency combs enable quantum-induced coherence to occur despite the substantially different dates of birth of the various pulses. {\it This is the key of the qCOMBPASS concept}.
In turn, this property is the reason why qCOMBPASS does not require a quantum memory to coherently store any photon. In some sense, the whole comb is the storage of information.

qCOMBPASS potentially allows high performance for quantum remote sensing 
because it is based on frequency combs with zero carrier envelope offset frequency, that have coherence times larger than 2,000 seconds \cite{Picque2019}. This represents an enhancement of more than 6 orders of magnitude in usable interrogation time of a target of interest with respect to schemes utilizing quantum memories with storage times on the order of milliseconds. For the same reasons,  
qCOMBPASS potentially unveils unprecedented functionality for quantum remote sensing 
because the large coherent times of frequency combs enable sensing ranges of more than 100's of kilometers, much larger than current state-of-the-start technology based on QI and quantum memories whose sensing range is a few meters, as demonstrated in tabletop experiments \cite{Barzanjeh2020,Zhang2013,Lopaeva2013,Barzanjeh2015,Gregory2020}. As discussed later in the paper, our prediction of record-high remote quantum sensing performance takes into account realistic parameters for distance-to-target, atmospheric absorption and turbulence, light diffraction, light squeezing amplitude, as well as photon collection efficiency and single-photon detection efficiencies with current technology based on superconducting nanowire single-photon detectors (SNSPDs). 

%%%%%%%%%%
\vspace{0.2cm}
\section{Perturbative qCOMBPASS}

For a more detailed description of qCOMBPASS, and to closely follow the original ZWM experiment, we first consider the perturbative regime of qCOMBPASS in which photon pairs are generated via SPDC with a low-power pump. In this regime, a TMSVc boils down to a biphoton quantum frequency comb \cite{Reimer2016,Kues2017,Guo2017,Pasquazzi2018,Guidry2022}, i.e., a train of periodic SPDC signal/idler pulse pairs containing one photon per pulse. A classical frequency comb pumps a nonlinear crystal 
(e.g., periodically poled lithium niobate PPLN) and generates the biphoton comb with signal  $\omega_{s,m}$  and idler $\omega_{i,m}$ comb frequencies satisfying energy-momentum conservation (phase matching conditions, $m$ is the comb index). To illustrate the qCOMBPASS concept, we consider a two-dimensional model, use the retroreflection version of ZWM, and describe the spatiotemporal dynamics stroboscopically (Fig. \ref{Fig1}). 

It is convenient to discretize space and time as  $x_j=j \Delta$ and $t_k=k t_{rep}$, label pulses by their horizontal position and direction of motion at each stroboscopic step, and associate the vertical position $y_{p,s,i}$ to pump, signal, and idler pulses, respectively. ($t_{rep}$ is the repetition period of the pump comb and $\Delta=c t_{rep}$ is the distance between consecutive pulses.) Every pump pulse generates a signal/idler pair with probability amplitude $g\ll 1$, that for simplicity is assumed to be the same for all comb frequencies. A first pump pulse moving to the right enters the nonlinear crystal at position $x_0$ and generates a right-moving signal/idler photon pair. Suppose that at time $t_0$ a right-moving pump pulse enters the crystal and generates an SPDC pair.
In the no-depletion pump approximation 
(pump quantum state is minimally perturbed by the SPDC process of converting one pump photon into signal/idler photon pairs), 
the state at time $t_0$ is
\begin{equation}
|\psi(t_0) \rangle =\frac{(1+ g |s^{(0)} \rangle_{0}^R |i^{(0)} \rangle_{0}^R )\; |p\rangle_{0}^R
}{\sqrt{1+g^2}}
\end{equation}
The notation is: $|s/i^{(j)} \rangle^{\alpha}_k$, where $j$ is the time of birth $t_j$ of the single-photon signal/idler pulse, $k$ denotes the pulse current position $x_k$, and $\alpha=R/L$ the direction of motion. For the pump pulses we do not indicate a birth time. 
Note that, in contrast to the signal and idler pulses that contain one photon each, the pump pulse is a frequency comb laser with a large amount of photons per pulse.
At a time $t_1$ the pulses move to the right and a new pump pulse enters the crystal and generates a new signal-idler pair:
\begin{equation}
|\psi (t_1) \rangle =  \frac{(1+ g |s^{(1)}\rangle_{0}^R  |i^{(1)}\rangle_{0}^R ) |p\rangle_{0}^R}
{\sqrt{1+g^2}} 
\odot
\frac{(1+ g |s^{(0)}\rangle_{1}^R |i^{(0)}\rangle_{1}^R ) |p\rangle_{1}^R}
{\sqrt{1+g^2}}
\end{equation}
We use $\odot$ to separate signal/idler pulse pairs according to their time of birth.
At time $t_2$ the first signal pulse continues to the target but the first pump/idler pulses retro-reflect and change direction of propagation:
\begin{equation}
|\psi (t_2) \rangle = 
\frac{(1+ g |s^{(2)}\rangle_{0}^R |i^{(2)} \rangle_{0}^R ) |p\rangle_{0}^R}
{\sqrt{1+g^2}}  
\odot
\frac{(1+ g |s^{(1)}\rangle_{1}^R |i^{(1)}\rangle_{1}^R ) |p\rangle_{1}^R}
{\sqrt{1+g^2}} 
\odot
\frac{(1+ g |s^{(0)}\rangle_{2}^R |i^{(0)} \rangle_{1}^L ) |p\rangle_{1}^L}
{\sqrt{1+g^2}}
\end{equation}
At time $t_3$ the retro-reflected pump pulse re-enters the crystal and generates a new left-moving signal-idler pair $(S,I)$: 
\begin{eqnarray}
|\psi (t_3) \rangle &=& 
\frac{(1+ g |s^{(3)}\rangle_{0}^R |i^{(3)}\rangle_{0}^R ) |p\rangle_{0}^R}
{\sqrt{1+g^2}}  
\odot
\frac{(1+ g |s^{(2)}\rangle_{1}^R |i^{(2)}\rangle_{1}^R ) |p\rangle_{1}^R} 
{\sqrt{1+g^2}}
\odot
\frac{(1+ g |s^{(1)}\rangle_{2}^R |i^{(1)}\rangle_{1}^L ) |p\rangle_{1}^L}
{\sqrt{1+g^2}}
 \nonumber \\
&\odot&
\frac{(1+ g |s^{(0)}\rangle_{3}^R |i^{(0)}\rangle_{0}^L )}
{\sqrt{1+g^2}} 
\odot
\frac{(1+ g |S^{(3)}\rangle_{0}^L  |I^{(3)} \rangle_{0}^L ) |p\rangle_{0}^L}
{\sqrt{1+g^2}}
\end{eqnarray}
At time $t_4$ the first left-moving idler pulses $(i, I)$ exit the crystal and are aligned in a Y-junction along the same path, and  the left-moving signal pulse $S$ follows another parallel path that will be aligned with signal pulses reflected off the target in the future:
\begin{eqnarray}
|\psi (t_4) \rangle &\mapsto&
\frac{(1+ g |s^{(4)}\rangle_{0}^R |i^{(4)}\rangle_{0}^R ) |p\rangle_{0}^R}
{\sqrt{1+g^2}}  
\odot 
\frac{(1+ g |s^{(3)} \rangle_{1}^R |i^{(3)}\rangle_{1}^R ) |p\rangle_{1}^R }
{\sqrt{1+g^2}}
\odot
\frac{(1+ g |s^{(2)} \rangle_{2}^R |i^{(2)} \rangle_{1}^L ) |p\rangle_{1}^L}
{\sqrt{1+g^2}}
\nonumber \\
&\odot &
\frac{(1+ g |s^{(1)} \rangle_{3}^R |i^{(1)} \rangle_{0}^L )  
(1+ g |S^{(4)} \rangle_{0}^L  |I^{(4)} \rangle_{0}^L ) |p\rangle_{0}^L}
{\sqrt{1+2g^2+g^4}}
\nonumber \\
&\odot &
\frac{(1+ g |s^{(0)} \rangle_{4}^R |{\cal I}^{(0)} \rangle_{-1}^L ) 
\odot
(1+ g |{\cal S}^{(3)} \rangle_{-1}^L | {\cal I}^{(3)}  \rangle_{-1}^L ) |p\rangle_{-1}^L}
{\sqrt{1+2g^2+g^4}}
\end{eqnarray}
where the symbol $\mapsto$ means path identity is applied. 
At time $t_5$, 
\begin{eqnarray}
|\psi (t_5) \rangle &\mapsto&
\frac{(1+ g |s^{(5)} \rangle_{0}^R |i^{(5)}\rangle_{0}^R ) |p\rangle_{0}^R}
{\sqrt{1+g^2}}    
\odot 
\frac{(1+ g |s^{(4)}\rangle_{1}^R |i^{(4)} \rangle_{1}^R ) |p\rangle_{1}^R} 
{\sqrt{1+g^2}}  
\odot
\frac{(1+ g |s^{(3)} \rangle_{2}^R |i^{(3)} \rangle_{1}^L ) |p\rangle_{1}^L}
{\sqrt{1+g^2}}  
\nonumber \\
&\odot &
\frac{(1+ g |s^{(2)} \rangle_{3}^R |i^{(2)} \rangle_{0}^L )  (1+ g |S^{(5)} \rangle_{0}^L |I^{(5)} \rangle_{0}^L ) |p\rangle_{0}^L}
{\sqrt{1+2g^2+g^4}}
\nonumber \\
&\odot&  
\frac{(1+ g |s^{(1)} \rangle_{4}^R |{\cal I}^{(1)} \rangle_{-1}^L )  (1+ g |{\cal S}^{(4)}\rangle_{-1}^L |{\cal I}^{(4)}\rangle_{-1}^L ) 
|p\rangle_{-1}^L}
{\sqrt{1+2g^2+g^4}}
\nonumber \\
&\odot&
\frac{(1+ g |s^{(0)} \rangle_{5}^R |{\cal I}^{(0)}\rangle_{-2}^L ) 
\odot
(1+ g |{\cal S}^{(3)}\rangle_{-2}^L |{\cal I}^{(3)}\rangle_{-2}^L ) |p\rangle_{-2}^L}
{\sqrt{1+2g^2+g^4}}
=
|\chi \rangle_{\neq -2} \odot |\chi \rangle_{-2}
\label{s6}
\end{eqnarray}
where $|\chi \rangle_{\neq -2}$ involves pulses not at the pre-detection position, 
and $|\chi \rangle_{-2}$ includes pulses at $x=-2$.
We assume left-moving idler photons are measured at $x_{-2}$ via a projective measurement $\Pi_{{\cal I},-2}^L$ on number states of idler photons. After path identity,
\begin{eqnarray}
&& |\chi \rangle_{-2} \mapsto
 \frac{ [\; 1  + g |s^{(0)} \rangle_{5}^R |{\cal I}\rangle_{-2}^L  + 
g |{\cal S}\rangle_{-2}^L | {\cal I}\rangle_{-2}^L
+ g^2 |s^{(0)} \rangle_{5}^R  |{\cal S}\rangle_{-2}^L  |2{\cal I}\rangle_{-2}^L \; ]  \;  |p\rangle_{-2}^L}
{\sqrt{1+2g^2+g^4}}
\label{s7}
\end{eqnarray}
where $|2{\cal I}\rangle_{-2}^L$ is a two-photon Fock state in the idler line after path identity.
Note we have dropped the time stamps (date of birth) of idler photons because what matters is that they are all indistinguishable and located at position $x_{-2}$, ready to be measured. The time stamp of when they were born is no longer relevant.
From all possible measurement outcomes, we post-select measurement results in which the idler detector clicks
(i.e., $1, 2, 3, \ldots $ idler photons), as these are the only ones useful for sensing. Hence, 
\begin{equation}
\Pi_{{\cal I},-2}^L= |{\cal I} \rangle \langle {\cal I} |_{-2}^L + |2{\cal I} \rangle \langle 2{\cal I} |_{-2}^L  
+ |3{\cal I} \rangle \langle 3{\cal I} |_{-2}^L +
\ldots
\end{equation}
The state after measurement is decribed by the density matrix 
\begin{eqnarray}
\rho(t_5) &=&{\cal N}^{-1} {\rm Tr}_{{\cal I}} [\Pi_{{\cal I},-2}^L  ( |\chi \rangle_{\neq -2} \odot |\chi \rangle_{-2}) 
(\langle \chi|_{\neq -2} \odot \langle \chi|_{-2})  
\otimes \rho_E  \otimes \rho_T \otimes \rho_B \Pi_{{\cal I},-2}^L]
\nonumber \\
&=&
{\cal N}^{-1} {\rm Tr}_{{\cal I}} [\Pi_{{\cal I},-2}^L  |\chi \rangle \langle \chi|_{-2}  \Pi_{{\cal I},-2}^L ] 
\odot  |\chi \rangle \langle \chi|_{\neq -2}  
\otimes \rho_E  \otimes \rho_T \otimes \rho_B 
\end{eqnarray}
where 
\begin{eqnarray}
 {\rm Tr}_{{\cal I}} [\Pi_{{\cal I},-2}^L  |\chi \rangle \langle \chi|_{-2}  \Pi_{{\cal I},-2}^L ] 
&=&
\left\{ g^2 |s^{(0)}\rangle \langle s^{(0)}|_{5}^R   + g^2 |{\cal S}\rangle \langle {\cal S}|_{-2}^L    
+
g^2  \left[
|s^{(0)}\rangle_{5}^R \langle {\cal S}|_{-2}^L + |{\cal S}\rangle_{-2}^L  \langle s^{(0)}|_{5}^R \right] 
\right.
\nonumber \\
&+&
\left.
g^4 |s^{(0)} \rangle \langle s^{(0)}|_{5}^R  \;  |{\cal S}\rangle \langle  {\cal S}|_{-2}^L 
\; \right\} \;  |p\rangle \langle p|_{-2}^L
\label{s10}
\end{eqnarray}
and the normalization is 
\begin{eqnarray}
{\cal N} &=&
{\rm Tr}_{all} 
[\Pi_{{\cal I},-2}^L  |\chi \rangle \langle \chi|_{-2}  \Pi_{{\cal I},-2}^L  \odot  |\chi \rangle \langle \chi|_{\neq -2}  
\otimes \rho_E  \otimes \rho_T \otimes \rho_{B} ]
\end{eqnarray}
${\rm Tr}_{all}={\rm Tr}_{E,T,B, s,i,S,l,{\cal S},{\cal I}}$ and 
$\rho_E, \rho_T, \rho_B $ describe the states of the environment,
the target, and the absorbing block for pump and signal pulses. Left-moving signal photons are not measured. 
At time $t_6$, left-moving pump and signal pulses are absorbed in a block via a unitary operation:
$U_{B} |B\rangle \;  |{\cal S}\rangle_{-2}^L \; |p\rangle_{-2}^L = |B'\rangle \;  |0_{\cal S} \rangle_{-3}^L \; 
|0_p \rangle_{-3}^L$, where $ |B\rangle$ is the state of the block prior to absorption and $|B'\rangle$ post-absorption.
Hence, from Eq. (\ref{s10}),
\begin{eqnarray}
&& U_{B} {\rm Tr}_{{\cal I}} [\Pi_{{\cal I},-2}^L  |\chi \rangle \langle \chi|_{-2}  \Pi_{{\cal I},-2}^L ]  \otimes \rho_B U^{\dagger}_{B} 
=g^2 |s^{(0)}\rangle \langle s^{(0)}|_{6}^R   |B\rangle \langle B|+ g^2 |B'\rangle \langle B'|    
\nonumber \\
&&+
g^4 |s^{(0)} \rangle \langle s^{(0)}|_{6}^R  \; |B'\rangle \langle B'|
+
g^2  \left[ |s^{(0)}\rangle_{6}^R   \langle B'| + |B' \rangle \langle s^{(0)}|_{6}^R    \right] 
\end{eqnarray}
The first signal probe pulse $|s^{(0)} \rangle^R_{5}$ is now in $|s^{(0)}\rangle^R_{6}$ and continues to target. 
Diagonal terms $|s^{(0)}\rangle \langle s^{(0)}|_{6}^R$ will not contain  phase information, but
terms $|s^{(0)}\rangle_{6}^R   \langle B'| + |B' \rangle \langle s^{(0)}|_{6}^R$ will.

The stroboscopic process continues {\it mutatis mutandis} until the first signal probe photon reflects off the target at position $x_T$ at time $t_T$. 
We model this process as usual in quantum optics via an effective beam splitter unitary operator $\mathbb{U}_{T}$ that transforms pulse state 
$|s \rangle $ into the state $ \sqrt{\kappa} |s_r \rangle  + \sqrt{1-\kappa} |s_{as}\rangle$, where 
$|s_r \rangle$ corresponds to the reflected pulse, 
$|s_{as}\rangle $ to an absorbed/scattered pulse in the target,
and $\sqrt{\kappa}=r e^{\phi_r}$ is the complex reflection coefficient of the target.
Information about the target is imprinted on the state of the reflected pulse: $|s_r \rangle_T^L=\int d\omega \tilde{\eta}(\omega) \sqrt{\kappa(\omega)} |s_{\omega} \rangle_{T}^L$, with $\tilde{\eta}(\omega)$
characterizing the spectrum of the pulse. 
The first pulse returns to the qCOMBPASS station at time $t_{2T}$,
is delayed by one repetition period in a delay line (see Fig. 1), makes a reverse pass through the crystal at time $t_{2T+2}$, and is at the pre-detection position $x_{-1}$ at time $t_{2T+3}$. 
We only write the  ``minimal" part of the whole state of the combs that is relevant for the detection process at pre-detection position $x=-1$ and will provide phase information:
\begin{eqnarray}
\rho (t_{2T+3}) &\cong&
g^2\sum_m   
\eta^2_m  
|{\cal S}_m\rangle \langle {\cal S}_m| \otimes  |{\cal I}_m\rangle \langle {\cal I}_m| 
+
g^2\sum_m   
 \eta^2_m  
|{\cal I}_m\rangle \langle {\cal I }_m| \otimes  
|s_m^{(2T-1)} \rangle^R_4  \langle s_m^{(2T-1)}|^R_4
\nonumber \\
&+ &
g^3 \sum_m \eta^3_m \sqrt{\kappa}  \;   
| {\cal S} _m \rangle   \langle  {\cal S} _m| \otimes
|{\cal I}_m \rangle  \langle  {\cal I} _m| \otimes
|s_m^{(2T-1)} \rangle^R_4 \langle 0|  
\nonumber \\
&+ &
g^3 \sum_m \eta^3_m \sqrt{\kappa^*}  \;   
| {\cal S} _m \rangle   \langle  {\cal S} _m| \otimes
|{\cal I}_m \rangle  \langle  {\cal I} _m| \otimes
|0 \rangle \langle s_m^{(2T-1)}| ^R_4  
+ O(g^4) 
\label{s13}
\end{eqnarray}
where the sums are over comb indices $-M \le m \le M$ and we discard non-diagonal terms in frequency because they cannot produce ZWM interference. We assume each pulse has the Gaussian form $E(t)=E_0e^{-t^2/\tau^2}$ and therefore a spectrum 
$\tilde{E}(\omega)$ proportional to $e^{-\omega^2\tau^2/4}$. The spectral weights $\eta_m$ are then defined as
\begin{equation}
\eta_m=\frac{e^{-m^2 \omega^2_{rep} \tau^2/4}}{\sum_{m=1}^M e^{- m^2 \omega^2_{rep} \tau^2/4}}
\end{equation}
The first term in Eq. (\ref{s13}) is quadratic in $g$ and corresponds to a signal-idler pair of pulses that was generated in a reverse pass by a pump pulse. 
The second term is also quadratic in $g$ and encompasses and idler pulse at the pre-detection position and a right-moving signal pulse traveling to target. 
The third and fourth terms are cubic in $g$ and correspond to the ZWM path-identity interference term
between the signal/idler reverse pass and the reflected signal pulse/retro-reflected idler pulse. 

At the pre-detection time-bin there are two left-moving idler pulses $(i,I)$ sharing the same mode at position $x_{-1}$, namely a 
newborn idler pulse generated in a forward pass that retro-reflected and a ``younger"  newborn idler pulse generated in a reverse pass. Also, two left-moving signal pulses 
$(s_r,S)$ share another path at the same position $x_{-1}$: the old probe signal pulse that reflected off the target and the newest signal pulse. Note that the twin signal pulse born together with $i$ is at current position $x_4$ and on the way to the target, whereas the twin idler pulse born together with $s_r$ was measured by the idler detector long in the past. After the first signal pulse returns from target, qCOMBPASS enters a stationary regime.

Path identity enables the identification of left-moving signal pulse modes at the input of one Y-junction with a single signal mode at the output \cite{Chiao1985}: 
$s_r,  S \mapsto {\cal S}$. The same holds for the left-moving idler pulse modes 
$i, I \mapsto {\cal I}$ at the other Y-junction.
Prior to measurement by the idler detector, the state of the system after path identity is a superposition of various
possibilities. To lowest order in SPDC perturbation theory, it contains the states of: 
a) the pump pulse (the leading contribution);
b) the signal/idler pulses generated in a reverse pass ($S, I \mapsto {\cal S}, {\cal I}$);
c) the retro-reflected idler pulse with its twin propagating to the target ($i, s \mapsto {\cal I}, s^R_4$); and
d) the reflected signal pulse $s_r$ together its twin idler that was detected long in the past ($s_r \mapsto {\cal S})$. Interference occurs 
between $S, I$ and $s_r, i$. We write a  ``minimal" form of the full quantum state (as indicated by $\cong$) in which
we show only contributions at the pre-measurement position $x_{-1}$ in which 
there is at least one photon in the idler line (idler detector will click).
We consider only equal comb frequencies, as unequal comb frequencies do not interfere in qCOMBPASS, and we ignore pulses at positions other than $x_{-1}$ and pulse $s^R_4$, as they are irrelevant at the current detection time bin. Omitting pump pulses in the comb sums, we have the density matrix 
\begin{eqnarray}
\rho &\cong&
g^2 \sum_m    \eta^2_m  |{\cal S}_m\rangle \langle {\cal S }_m| \otimes  |{\cal I}_m\rangle \langle {\cal I}_m| 
+
g^2 \sum_m   
\eta^2_m |{\cal I}_m\rangle \langle {\cal I}_m| \otimes  |s_m\rangle \langle s_m|^{R}_{4} 
\nonumber \\
&+&
g^3 \sum_m    \eta^3_m \sqrt{\kappa_m}  \;   
| {\cal S} _m \rangle   \langle  {\cal S} _m| \otimes
|{\cal I}_m \rangle  \langle  {\cal I} _m| \otimes
|s_m\rangle^R_4 \langle 0|^R_4 
\nonumber \\
&+&
 g^3 \sum_m    \eta^3_m \sqrt{\kappa^*_m}  \;   
| {\cal S} _m \rangle   \langle  {\cal S} _m| \otimes
|{\cal I}_m \rangle  \langle  {\cal I} _m| \otimes
|0 \rangle^R_4 \langle s_m| ^R_4  
\label{eq2}
\end{eqnarray}
 
 Assuming the idler detector has frequency-resolution (e.g., a comb spectrometer), the photo-counts at the idler comb frequency $\omega_{i,m}$ are
\begin{eqnarray}
N_m(t_{2T+3}) &=&{\rm Tr}_{all} [a^{\dagger}_{{\cal I},m} a_{{\cal I},m} \rho(t_{2T+3}) ]
=2g^2 \eta^2_m + O(g^4)
\end{eqnarray}
The terms cubic in $g$ are null when taking the trace because of the non-diagonal factors
$|s_m^{(2T-1)} \rangle^{R}_{4} \langle 0|$ and $|0 \rangle \langle s_m^{(2T-1)}|^{R}_{4}$. 
As a result, ZWM interference does not occur and target information is lost. 

In a usual ZWM experiment with continuous wave (cw) light, the state after path identity is described by 
$\rho_{ZWM} = 2g^2 (1+ {\rm Re} \sqrt{\kappa})  |{\cal S} \rangle \langle{\cal S}| \;  |{\cal I} \rangle \langle{\cal I}|$
and the photo-count at the idler detector is
$N_{{\cal I}}={\rm Tr} [ a^{\dagger}_{\cal I} a_{\cal I} \rho_{ZWM}]= 2g^2  (1+ {\rm Re} \sqrt{\kappa})$. The second term 
contains reflection amplitude and phase information of the target and stems from interference after path identity between the reverse signal/idler pulses and the retro-reflected
signal/idler pulses.
In qCOMBPASS, this interference should be contained in the last two terms of $\rho$ that are proportional to $g^3$. Note
that the difference in power-law is due to the fact that qCOMBPASS  involves three SPDC interactions while ZWM involves only two. But ZWM requires a quantum memory while qCOMBPASS does not.
Standard ZWM  involves two pairs of twin signal-idler photons, and qCOMBPASS involves one pair, plus an idler and a signal, and the last two are not twins. 

%%%%
\section{qCOMBPASS chopper protocol}

The chopper protocol solves the problem of absence of interference.
In the implementation of qCOMBPASS described so far, the computation of photo-counts in the idler detector using $\rho$ as written in Eq.\,(\ref{eq2}) gives no ZWM interference because $ |0\rangle^{R}_{4}$ and $|s_m \rangle^{R}_{4}$ are orthogonal:  photo-counting at the idler detector does not contain target information. Although the signal pulse $|s\rangle_{4}^R$
is not at the detection location, it cannot be dismissed in the evaluation of photo-counts because it is entangled with its idler twin about to be detected.
qCOMBPASS solves this lack of interference by introducing a  ``chopper protocol". In its simplest form, the chopper alternatively blocks every other right-moving signal pulse going to the target, so that when the forward idler pulse retro-reflects and arrives at position $x_{-1}$, its twin signal pulse was already chopped. We model the action of the chopper on the mode $s$ via a unitary operation $\mathbb{U}_{chop}$
that absorbs all photons in the forward signal pulse and accordingly changes the state of the chopper from $|C\rangle$ to $|C'\rangle$. Chopping doubles the period of the signal probe comb and halves the separation between its comb teeth, hence the reflected comb has comb frequencies
$\omega_{s_r,m'}=\omega_c+ m' \omega_{rep}/2$. After path identity,
\begin{eqnarray}
\rho &\cong&
g^2 \sum_m   
 \eta^2_m  |{\cal S}_m\rangle \langle {\cal S }_m| \otimes  |{\cal I}_m\rangle \langle {\cal I}_m| 
\otimes |C\rangle \langle C| 
+
g^2 \sum_m \eta^2_m |{\cal I}_m\rangle \langle {\cal I}_m| \otimes  |C'\rangle \langle C|
\nonumber \\
&+ &
g^3 \sum_m \sum_{m'}  \eta^2_m \eta_{m'}  \sqrt{\kappa_{m'}}  \;   
| {\cal S} _{m'} \rangle   \langle  {\cal S} _{m'}| \otimes
|{\cal I}_m \rangle  \langle  {\cal I} _m| 
\otimes |C' \rangle \langle C| 
\nonumber \\
&+ &
g^3 \sum_m \sum_{m'}  \eta^2_m \eta_{m'}  \sqrt{\kappa^*_{m'}}  \;   
| {\cal S} _{m'} \rangle   \langle  {\cal S} _{m'}| \otimes
|{\cal I}_m \rangle  \langle  {\cal I} _m| 
\otimes |C \rangle \langle C'|  
\end{eqnarray}
Interference is now possible because the macroscopic chopper is 
weakly perturbed by few-photon absorption and hence $\langle C | C' \rangle \approx 1$. Also, comb teeth corresponding to the reverse comb
$|S\rangle$ can interfere with even comb teeth of $|s_r\rangle$, i.e., when $m'=2m$.  After path identity, ZWM quantum-induced coherence at the idler detector occurs when the chopper blocks a new signal pulse from going to the target, and it does not occur when it allows it to go to the target. The perturbative qCOMBPASS photo-count at 
idler frequency $m$ is then
\begin{eqnarray}
N_{{\cal I},m}(g\ll 1; {\rm chopper \; OFF})  &=& 
2 \mu_d g^2 \eta^2_m   + O(g^4)
 \nonumber \\
N_{{\cal I},m}(g\ll 1; {\rm chopper \; ON})  &=& 2 \mu_d g^2 \eta^2_m
+ 4    O_{{\cal S}} O_{{\cal I}} \mu_d
g^3 \eta^3_m  r_{m}\cos\Psi_m + O(g^4)
\label{spdc}
\end{eqnarray}
where $\mu_d$ is the detector efficiency.
Photo-counts measured at the idler comb frequency $\omega_{i,m}$ when the chopper is in the ON-state contain target information at the signal comb frequency $\omega_{s,m}=\omega_{s_r,2m}$ encapsulated in the complex reflection coefficient of the target,
$\sqrt{\kappa_{m}}=r_{m} e^{i\phi_{r,m}}$.
$O_{{\cal I},m}$ and $O_{{\cal S},m}$ are the spatiotemporal overlaps of pulses $(I_m,i_m)$ and
$(S_m,s_{r,2m})$ after path identity, respectively (see Appendix). 
$\Psi_m=\varphi_{s,m}(d) + \varphi_{i,m} 
- \varphi_{S,m}-\varphi_{I,m}+\phi_{r,m}$
where the 
$\varphi$'s are propagation phases of the $s, i,S,I$ pulses at comb frequencies indicated by the frequency subscripts. The signal propagation phase depends on the distance to the target $d$ via the phase
$\varphi_{s,m}(d)= 2 \omega_{s,m} d/c$.
Since maximal mechanical chopping frequencies are much slower than comb 
repetition frequencies (10s kHz vs $>$100s MHz), it is not possible to chop every other signal pulse as discussed above. Instead, one can block a train of  $p$ consecutive pulses, then unblock the next train of $p$ pulses, etc., and the photo-count formula is the same as Eqs. (\ref{spdc}) but with the reflection coefficient evaluated at the comb tooth of the chopped probe comb that matches the comb tooth of the detected idler comb.
Piezo-delay stages can be used to tune the qCOMBPASS interference pattern, e.g., 
a movable mirror on the path of pulses $i$ enables the buildup of 
the pattern by rastering the propagation phase 
$\varphi_{i,m}$ and controlling the overlap 
$O_{{\cal I},m}=e^{-(t-\tau_{i,I})^2/\sigma^2_t}$ via time-delay $\tau_{i,I}$.
The  visibility of the perturbative qCOMBPASS interference pattern is 
\begin{equation}
V_{{\cal I},m}=\frac{N^{(max)}_{{\cal I},m} - N^{(min)}_{{\cal I},m} }{ 
N^{(max)}_{{\cal I},m} + N^{(min)}_{{\cal I},m} }=
2  O_{{\cal S}} O_{{\cal I}} g \eta_{m} r_{m}
\end{equation}
smaller than standard ZWM by a factor $2g \eta_m$ and much smaller than non-perturbative ZWM \cite{Volkoff2023}.
Eqs.\,(\ref{spdc}) assume no photon loss/noise and neglect
electric field fall-off with distance, so that they are only applicable to near-range sensing. In this regime, ZWM (with either cw or pulsed light and with either small or large $g$) outperforms 
qCOMBPASS because it can employ a quantum memory. However, for quantum remote sensing this may be impossible because of the long memory times required, and it is precisely in this remote-range regime where qCOMBPASS is most advantageous, as discussed next.

%%%%
\vspace{0.2cm}
\section{Non-perturbative qCOMBPASS} 

Perturbative qCOMBPASS as just described suffers from low conversion efficiency and is highly susceptible to photon loss/noise because each probe pulse contains only one photon with low probability. This shortcoming is overcome in non-perturbative qCOMBPASS based on squeezed light. Quadrature squeezing levels of an electromagnetic field have reached $\sim 15$ dB with current technology \cite{Vahlbruch2016} in applications ranging from gravitational wave detection \cite{Tse2019} to continuous-variable one-way quantum communications \cite{Pogorzalek2019}, and a qCOMBPASS transceiver based on two-mode squeezed vacuum quantum frequency combs (TMSVc) can likewise be beneficial for quantum remote sensing.
A TMSVc is a train of periodic identical pairs of pulses in a two-mode squeezed vacuum state
$\mathbb{S}_{a,b} |0\rangle$. The operator 
\begin{equation}
\mathbb{S}_{a,b}=\exp(\zeta^* a^{\dagger} b^{\dagger}- \zeta a b)
\end{equation}
squeezes the quantum vacuum with a (complex) strength $\zeta=g e^{\phi_g}$; $a^{\dagger}$ and $b^{\dagger}$  create photons in modes $a$ and $b$, respectively. 
The level of squeezing associated with this state is $-10 \log_{10} \exp(-2g)$ dB.
For simplicity we assume degenerate signal and idler modes.
TMSVc can be generated using SPDC combs or Kerr microcombs \cite{Yang2021}, can have large squeezing amplitudes, and can contain many photons per mode. Since the goal of qCOMBPASS is remote sensing, we must include the 1/{\boldmath$|x|$} far-field decay with distance of the electric field amplitude of the spatial mode of the signal probe pulses (see Section VIII). We must also account for effects on these pulses, such as loss and noise in a turbulent atmosphere, as they propagate to and from a distant target (see Sections VII and IX). 
We model the loss with a unitary beam splitter operator $\mathbb{U}_{E}$
that transforms $|s\rangle$ into $ \sqrt{\xi} |s\rangle + \sqrt{1-\xi} |s_E\rangle$. Here 
$|s_E\rangle$ corresponds to a lost pulse and $\sqrt{\xi}=\ell e^{i \phi_{\xi}}$ is the complex loss/noise parameter.  This is continuously applied to the probe pulse at every position, i.e., $\xi=\xi(x)$. As in the perturbative case, and for the same reason, non-perturbative qCOMBPASS requires the chopper protocol.

%%%%
\vspace{0.2cm}
\section{qCOMBPASS path identity}

Path identity between pulses $(S, s_r)$ is implemented in a beam combiner, and the same is done for pulses
$(I, i)$ in another beam combiner. In a free-space qCOMBPASS implementation, a 50-50 beam splitter serves as a
beam combiner, but part of the input is lost through the second port of the beam splitter, causing reduced efficiency. 
In an on-chip qCOMBPASS implementation, a Y-junction 
\cite{Chiao1985}
is the ideal beam combiner that transforms two input modes $a,b$ into a single  output mode $c$ via mode path identity 
$a,b \mapsto c$. As a simple example, one can take two coherent states 
$|\alpha \rangle_a=\mathbb{D}(\alpha) |0\rangle_a$ and  $|\beta \rangle_b=\mathbb{D}(\beta) |0\rangle_b$ as the inputs of a Y-junction.
The Y-junction implements a unitary operation that corresponds to the path identity operation.
The coherent state displacement operators are 
$\mathbb{D}(\alpha)=\exp(\alpha a^{\dagger} - \alpha^* a)$ and
$\mathbb{D}(\beta)=\exp(\beta b^{\dagger}- \beta^* b)$. Then, since $a^{\dagger}$ and $b^{\dagger}$ commute, 
\begin{eqnarray}
&& \mathbb{D}(\alpha) \mathbb{D}(\beta) |0\rangle_{ab}=
e^{\alpha a^{\dagger} + \beta b^{\dagger} - \alpha^* a - \beta^* b } |0\rangle_{ab}
\mapsto e^{ (\alpha+ \beta) c^{\dagger} - (\alpha+\beta)^* c} |0\rangle_c = \mathbb{D}(\alpha+\beta) |0\rangle_c
\nonumber \\
\end{eqnarray}
which is a coherent state of amplitude $\alpha+\beta$ in output port $c$, as expected. 
For TMSV states,
path identity via a Y-junction is enforced at the level of products of the squeezing operators $\mathbb{S}_{a,b}$. In qCOMBPASS we  implement path identity as follows:
\begin{eqnarray}
\!\!\! \!\!\! \!\!\! 
\mathbb{S}_{S,I} &\mapsto&  \mathbb{S}_{{\cal S},{\cal I}}=e^{\zeta^* {\cal S}  {\cal I}- h.c.} \nonumber \\
\mathbb{S}_{s_c,i} &\mapsto& \mathbb{S}_{s_c,{\cal I}}= e^{\zeta^* s_c  {\cal I}- h.c.}  \\
\mathbb{S}_{s_r,i_d} &\mapsto &
\mathbb{S}_{a_E,i_d} \cdot \mathbb{S}_{a_T,i_d} \cdot  \mathbb{S}_{{\cal S},i_d} 
=
e^{\zeta^* \sqrt{1-\xi} \; a_E  i_d - h.c.}
\cdot 
e^{\zeta^* \sqrt{\xi (1-\kappa)} \;  a_T i_d- h.c.}
\cdot 
e^{\zeta^* \sqrt{\kappa \xi} \; {\cal S} i_d -h.c.} 
\nonumber
\end{eqnarray}

Imperfect path identity can affect the qCOMBPASS transceiver and degrade interference. This effect can be modeled by replacing signal and idler operators as
$a^{\dagger}_{s} \rightarrow \sqrt{t_X} a^{\dagger}_{s}  + \sqrt{1-t_X} a^{\dagger}_{X}$ and
$a^{\dagger}_{i} \rightarrow \sqrt{t_Y} a^{\dagger}_{i}  + \sqrt{1-t_Y} a^{\dagger}_{Y}$. Here, $a^{\dagger}_{X,Y}$ are creation operators into dummy modes orthogonal to the signal-idler modes
and $t_{X,Y}$ are complex coefficients. Perfect alignment for path identity of signal and idler photons holds for $t_X=1$ and $t_Y=1$, respectively. 

%%%%
\vspace{0.2cm}
\section{qCOMBPASS transceiver state} 

The two-mode squeezing operator  
$\mathbb{S}_{a,b}$ is used to derive the density matrices of the various frequency combs employed in qCOMBPASS. The squeezing parameter $\zeta=g^{i \phi_g}$
is assumed to be frequency-independent. We write the  ``minimal" form of the state, as explained above.
The newest signal-idler pair of comb pulses generated by the reverse TMSVc comb has a density matrix
\begin{equation}
\rho_{S_m,I_m} \cong \mathbb{S}^{}_{S_m,I_m} |0\rangle \langle 0| \; \mathbb{S}^{\dagger}_{S_m,I_m}
\end{equation}
that corresponds to a two-mode Gaussian state. 
We assume a mechanical chopper with maximal chopping frequency 
$\sim 10$s kHz much slower than the pump comb repetition rate 
$\omega_{rep}/2 \pi >100$s MHz that chops train of $p$ consecutive pulses, then allows a next batch of $p$ pulses fly to target, then blocks the next batch of $p$ pulses, and so on. This causes the reflected signal comb to have larger period 
$p T$ and smaller tooth-to-tooth separation $\omega_{rep}/p$ than the original unchopped comb. In qCOMBPASS, a comb frequency $\omega_{s,m}$ of the signal comb $S$ can only undergo ZWM interference with reflected comb frequency 
$\omega_{s_r,pm}$ because
$\omega_{s_r,pm}=\omega_c + p m \omega_{rep}/p=
\omega_c + m \omega_{rep}=\omega_m$.
Note, however, that the chopping operation does not modify the comb frequencies of the twin idler comb.
The chopping operation results in retro-reflected new idler pulses with idler-chopper joint density matrix
\begin{equation}
\rho_{i_m,C} \cong 
\mathbb{U}_{chop} \; \mathbb{S}^{}_{s_{m},i_m} |0\rangle \langle 0| \otimes \rho_C  \; \mathbb{S}^{\dagger}_{s_{m},i_m} \mathbb{U}_{chop}
\end{equation}
where the chopping operation performs the unitary transformation
$ \mathbb{U}_{chop}  |s \rangle \otimes |C\rangle = |0\rangle \otimes |C'\rangle$ 
that absorbs all signal photons of the forward TMSVc when the chopper is ON, and is the identity operator when the chopper is OFF.
We use the fact that a macroscopic chopper is minimally perturbed by absorption of signal photons and has a stationary density matrix $\rho_C$, and we approximate
$\rho_{i_m,C}  \approx \rho_{i_m} \otimes \rho_C$,
where $\rho_{i}$ is a one-mode non-Gaussian state $i$ generated by photon subtraction from the squeezed vacuum. The three pulses $S,I,i$ do not leave the qCOMBPASS station and are unaffected by loss, noise, and diffraction, and acquire propagation phases $\varphi_{S}$, $\varphi_{I}$ and $\varphi_{i}$ prior to path identity. 

Old idler pulses corresponding to old twin signal probe pulses 
that went to target are measured well before the latter return to
the qCOMBPASS station. Idler detection generates a one-mode non-Gaussian state $\rho_{s}$
by photon subtraction of idler photons of  forward TMSVc via projective measurements onto a Fock basis.
The idler detector has frequency comb resolution but does not  have single-photon number resolution.
The post-measurement density matrix of the signal pulse before leaving the qCOMBPASS station to target is
\begin{equation}
\rho_{s_{m}} \cong 
\frac{ {\rm Tr}_{i}  [ \Pi_{i}  \mathbb{S}^{}_{s_{m},i_{m}} 
|0\rangle \langle 0| \mathbb{S}^{\dagger}_{s_{m},i_{m}} 
\Pi_{i} ]}
{{\rm Tr}_{s,i}  [  \Pi_{i} \mathbb{S}^{}_{s,i} 
|0\rangle \langle 0|
\mathbb{S}^{\dagger}_{s,i}]} 
\end{equation} 
where the denominator is the normalization factor ${\cal N}_s$. Because $p$ consecutive pulses are blocked and unblocked from
the probe signal comb, its comb teeth are at frequencies
$\omega_{s_r,m'}=\omega_c/2 + m' \omega_{rep}/p$ , and only teeth with $m'=p m$ can have ZWM interference with the local combs
$S,I, i$. The signal-environment-target joint density matrix is
\begin{equation}
\rho_{s_{r,m},E,T} 
\cong
 \mathbb{U}_{E_{m}} \mathbb{U}_{T_{m}} \;  \rho_{s,m} 
\otimes \rho_E \otimes \rho_T \;
\mathbb{U}_{T_{m}}^{\dagger} \mathbb{U}_{E_{m}}^{\dagger}
\end{equation}
$\mathbb{U}_{E}$ is the unitary operator describing the interaction between the environment and the signal pulse going to the
target, and is modeled as an effective beam splitter. One input port is the signal, the other is vacuum. Output ports are the perturbed signal and the lost photon channel. The beam splitter unitary operator is
\begin{eqnarray}
\mathbb{U}_{E_{m}}(z) = 
\left( 
{\begin{array}{cc}
\sqrt{\xi_{m}(z)} & \sqrt{1-\xi_{m}(z)}  \\
-\sqrt{1-\xi_{m}(z)} & \sqrt{\xi_{m}(z)} \\
\end{array} } \right)
\end{eqnarray}
where $\sqrt{\xi_{m}(z)}=\ell_{m}(z) e^{\phi_{\xi_{m}}(z)}$ is a complex quantity that encapsulates loss and noise effects. 
We call $z$ the vertical coordinate of the signal probe pulse (called $x_j$ in Fig. 1 of the main text).
The beam splitter has the pulse $|s\rangle$ and vacuum in the two input ports, and the transmitted pulse and an absorbed pulse 
$|s_E\rangle$ in the output ports. The unitary operation implements the mode conversion
$|s\rangle \rightarrow \sqrt{\xi(z)} |s\rangle +  \sqrt{1-\xi(z)}  |s_E\rangle$. 
Note that this unitary operation is continuously applied as the probe pulse travels to the target and depends on position through, e.g., the variation of the column density of the atmosphere.
$\mathbb{U}_{T}$ is the unitary operator for the signal-target interaction and is also modeled as a beam splitter:
\begin{eqnarray}
\mathbb{U}_{T_{m}} = 
\left( 
{\begin{array}{cc}
\sqrt{\kappa_{m}} & \sqrt{1-\kappa_{m}}  \\
-\sqrt{1-\kappa_{m}} & \sqrt{\kappa_{m}} \\
\end{array} } \right)
\end{eqnarray}
where $\sqrt{\kappa_{m}}= r_m e^{i \phi_{r,m}}$ is the complex reflection coefficient of the target. The beam splitter has the pulse $|s\rangle$ and vacuum in the two input ports, and the reflected pulse $|s_r\rangle$ and an absorbed/scattered pulse 
$|s_{as}\rangle$ in the target in the output ports. The unitary operation implements the mode conversion
$|s\rangle \rightarrow \sqrt{\kappa} |s_r\rangle +  \sqrt{1-\kappa}  | s_{as} \rangle$
at the location of the target.
An additional operation accounting for the interaction with the environment of reflected signal photons should be added,
 but it is easily shown that it is not needed to get the correct final result; a single 
unitary operation accounts for dissipation and noise processes in the whole round trip.
The unitary operator describing absorption of pump and signal pulses is defined similarly.
We assume the environment and  target are macroscopic and weakly perturbed: 
$\rho_{s_{r,m},E,T,B} \approx \rho_{s_{r,m}} \otimes \rho_E \otimes \rho_T \otimes \rho_B$.
The accumulated propagation phase of the probe pulse is
$\varphi_{s_{r,m}}=(2\omega_{s,m}d/c) \int_0^d n_{atm}(\omega_{s,m},z) dz$,  with $n_{atm}(\omega_{s,m},z)$ 
the height-dependent refractive index of the atmosphere. The accumulated loss/noise is contained in the complex quantity $\sqrt{\xi_{m}(d)}=\ell_{m} e^{i \phi_{\xi_{m}}}$, where
\begin{eqnarray}
\ell_{m} &=& \exp[- 2 \int_0^d {\mathscr L}(z,\omega_{s,m}) dz] \nonumber \\
\phi_{\xi_{m}} &= & \int_0^d {\mathscr N}(z,\omega_{s,m}) dz +  \int_d^0  {\mathscr N}'(z,\omega_{s,m}) dz
\end{eqnarray}
where 
${\mathscr L}(z,\omega_{s,m})$ accounts for dissipation and
${\mathscr N}(z,\omega_{s,m})$ for noise.
In the dissipation factor we assume equal attenuation to and from the target.
In the noise phase the two contributions are not necessarily equal because of random noise realizations. 
For an environment in thermal equilibrium, loss and noise functions are linked via the fluctuation-dissipation theorem $\langle\langle {\mathscr N}_{th}({\bf x},\omega) 
{\mathscr N}_{th}({\bf x}',\omega') \rangle\rangle=
\delta({\bf x}-{\bf x}') \delta(\omega-\omega') 2\hbar (1+n_{\beta}(\omega)) {\mathscr L}({\bf x},\omega)$, where $n_{\beta}(\omega)$ is the Bose thermal distribution and $\langle\langle (\ldots) \rangle\rangle$ denotes an
average over noise realizations. For a turbulent environment, equilibrium fluctuation-dissipation relations do not hold and a more complex analysis based on Kolmogorov theory 
is required \cite{Goodman1985}.
When the environment is the atmosphere, the attenuation corresponds to the height-dependent atmospheric extinction $\alpha_{atm}(x,\omega)$.

We then get the form of the qCOMBPASS transceiver state prior to path identity as
$\rho_{{\rm qCOMBPASS}}=(\rho_{{\rm qCOMBPASS},m})^{\otimes m}$, where
\begin{eqnarray}
\rho_{{\rm qCOMBPASS},m} &\cong&
\frac{1}{{\cal N}_s} \;  \mathbb{U}_B \mathbb{U}_{T,m} \mathbb{U}_{E,m}  \Pi_{i} \mathbb{S}^{}_{s_m,i} \; |0 \rangle \langle0|
 \otimes \rho_E \otimes \rho_T \otimes \rho_B\;  
(\mathbb{U}_B \mathbb{U}_{T,m} \mathbb{U}_{E,m}  \Pi_{i} \mathbb{S}_{s_m,i})^{\dagger} \nonumber \\
&\otimes & 
[ \mathbb{U}_{chop} \mathbb{S}^{}_{s,i_m} \;
|0 \rangle \langle0 |  \otimes \rho_C  \; 
(\mathbb{U}_{chop} \mathbb{S}_{s,i_m})^{\dagger}]
\otimes 
\mathbb{S}_{S_m,I_m} |0 \rangle \langle0| \; \mathbb{S}^{\dagger}_{S_m,I_m}
\end{eqnarray}
Henceforth we omit the comb index $m$. Assuming macroscopic environment, target, and chopper, the qCOMBPASS transceiver state can be written before path identity as
$\rho \cong \rho_{s_r} \otimes \rho_{i} \otimes \rho_{S,I} \otimes \rho_{ETBC}$, where
$\rho_{s_r} \cong {\rm Tr}_{i} \left[
\Pi_{i} \mathbb{S}^{}_{s_r,i}\; |0 \rangle \langle 0|  \; 
\mathbb{S}^{\dagger}_{s_r,i} \Pi_{i}\right]$,
$\rho_{i} \cong \mathbb{U}_{chop} \; \mathbb{S}_{s,i}\; |0 \rangle \langle 0| \;
\mathbb{S}^{\dagger}_{s,i}\; \mathbb{U}^{\dagger}_{chop}$,
$\rho_{S,I} \cong \mathbb{S}^{}_{S,I}\;|0 \rangle \langle 0| \;\mathbb{S}^{\dagger}_{S,I}$, and
$\rho_{ETBC} =
\rho_E \otimes \rho_T \otimes \rho_B \otimes \rho_C$.

We now discuss the structure of the qCOMBPASS transceiver. 
In dual-Fock basis the two-mode squeezing operator is
\begin{equation}
\mathbb{S}_{a,b} |0\rangle = 
\cosh^{-1}(g) \sum_{n=0}^{\infty} e^{i  n \phi_g} \tanh^n(g) |n,n\rangle_{a,b}.
\end{equation}
Before path identity, 
the reverse TMSVc comb has density matrix
$\rho_{S,I}= \mathbb{S}^{}_{S,I} |0\rangle \langle 0| 
\mathbb{S}^{\dagger}_{S,I}$ and is a two-mode
Gaussian state. The idler-chopper joint density matrix is
$\rho_{i,C} =\mathbb{U}_{chop}  \mathbb{S}^{}_{s,i} 
|0 \rangle \langle0 |_{s,i}  \otimes \rho_C 
\mathbb{S}^{\dagger}_{s,i}  \mathbb{U}^{\dagger}_{chop}$.
For a macroscopic chopper, 
$\rho_{i,C} \approx \rho_i \otimes \rho_C$, where
$\rho_i$ is a one-mode, non-Gaussian state  generated by signal-photon 
$s_c$ subtraction from the squeezed vacuum.  
The post-measurement density matrix of the signal pulse of the forward TMSVc before going to target is
$\rho_{s_r,E,T,B} =
{\cal N}_s^{-1} {\rm Tr}_{i}  
[\Pi_{i}  \mathbb{S}^{}_{s_r,i}
|0 \rangle \langle0|_{s,i} \otimes \rho_E \otimes \rho_T  \otimes \rho_B \;
\mathbb{S}^{\dagger}_{s_r,i}
\Pi_{i}]$.
The $\Pi_{i}$ are projectors onto Fock states of photons measured by the idler detector, the trace is over the states of the detector, and 
${\cal N}_s={\rm Tr}_{all} 
[ \Pi_{i} \mathbb{S}^{}_{s,i} |0\rangle \langle 0| \otimes \rho_E \otimes \rho_T  \otimes \rho_B \;\mathbb{S}^{\dagger}_{s,i}]$ is a normalization. 
Interaction with the environment and target is implemented via the unitary operators
$\mathbb{U}_{E}$ and $\mathbb{U}_{T}$, both modeled as beam splitters.
For macroscopic environments and 
targets, $\rho_{s_r,E,T,B} \approx \rho_{s_r} \otimes \rho_E \otimes \rho_T \otimes \rho_B$,
where $\rho_{s_r}$ is a one-mode non-Gaussian state generated by measurement.

The density matrix of the qCOMBPASS transceiver for quantum remote  sensing is a product over comb frequencies, 
$\rho_{{\rm qCOMBPASS}}=(\rho_{{\rm qCOMBPASS}_m})^{\otimes m}$, and each factor is the product 
$\rho_{{{\rm qCOMBPASS}}_m} \cong \rho_{{\cal S}_m,E,T} \otimes
\rho_{{\cal I}_m,C} \otimes \rho_{{\cal S}_m,{\cal I}_m}$
of an old signal pulse reflected off the target ($s_r$), a new retro-reflected idler pulse ($i$),
both from the forward TMSVc, and a new signal/idler pulse pair $(S,I)$
from the reverse TMSVc. As before, we only write the minimal form of the state. Explicitly, 
\begin{eqnarray}
\rho_{{\cal S}_{m},E,T,B} &\cong &
\mathbb{U}_{coll} \mathbb{U}_{E} \mathbb{U}_{T} \mathbb{U}_{B}\; \frac{1}{{\cal N}_s}
{\rm Tr}_{i} 
\left[
\Pi_{i} \mathbb{S}^{}_{{\cal S}_m,i_{m}}\;
|0 \rangle \langle 0| \; 
\mathbb{S}^{\dagger}_{{\cal S}_m,i_{m}}
\Pi_{i}
\right] 
\rho_{E} \otimes \rho_{T} \otimes \rho_B \;
\mathbb{U}^{\dagger}_{B}  \mathbb{U}^{\dagger}_{T}  \mathbb{U}^{\dagger}_{E} \mathbb{U}^{\dagger}_{coll}
\nonumber \\
\rho_{{\cal I}_m,C} &\cong&
\mathbb{U}_{chop} \; \mathbb{S}^{}_{s,{\cal I}_m}\;
|0 \rangle \langle 0| \otimes \rho_{C}\;
\mathbb{S}^{\dagger}_{s,{\cal I}_m}
\; \mathbb{U}^{\dagger}_{chop}
\nonumber \\
\rho_{{\cal S}_m,{\cal I}_m} &\cong &
\mathbb{S}^{}_{{\cal S}_m,{\cal I}_m}\;
|0 \rangle \langle 0| \;
\mathbb{S}^{\dagger}_{{\cal S}_M,{\cal I}_m}
\label{transceiver} 
\end{eqnarray}
$\mathbb{U}_{coll}$ is the unitary operator describing the photon collection,  
$\rho_{E}, \rho_{T}, \rho_B$, and $\rho_{C}$ are the density matrices of the environment, target, block, and chopper, respectively, that are minimally perturbed by the comb. The first line  includes the projective measurement
$\Pi_{i}$ of the detected idler pulse, whose twin signal pulse is propagating to the target, and the interaction with the absorbing block, target, environment, and signal beam collector. ${\cal N}_s$
normalizes the post-measurement state.
This results in a distance-dependent, renormalized 
squeezing amplitude for the reflected signal pulses,
$\tilde{\zeta}=\zeta  \mu_{coll} \sqrt{\kappa \xi}  f(d)$, 
where $\mu_{coll}$ is the collection efficiency of reflected signal pulses and $f(d)$
depends on the diffraction of the signal photons and the type of illumination of the target. Assuming Gaussian beams and full illumination, $f(d)=z_0/d$, where $z_0$ is the Rayleigh range (see Section VIII).
The second line in Eqs. (\ref{transceiver}) contains the chopping operation of a signal pulse whose twin idler pulse is about to be measured. We assume the chopper blocks a train of $p$ consecutive pulses, then unblocks the next train of $p$ pulses, etc., resulting in a transform-limited reflected comb with inter-tooth separation $\omega_{rep}/p$.

\begin{figure}[t]
\includegraphics[width=0.8\linewidth]{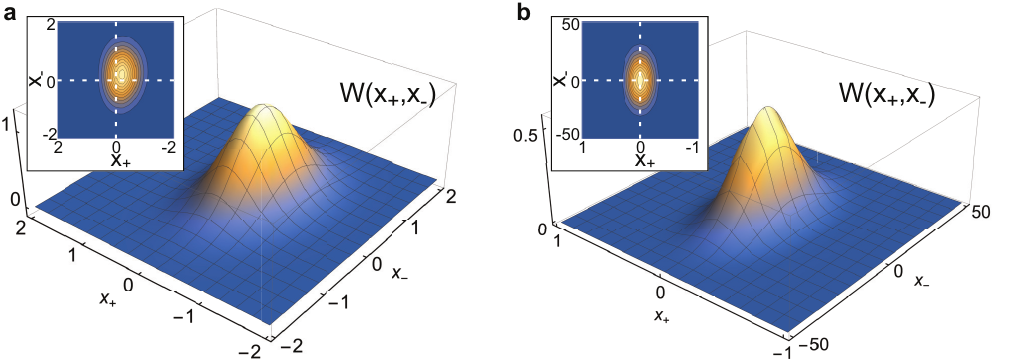}
\caption{{\bf qCOMBPASS Wigner function.}
Comparison between the Wigner functions of the qCOMBPASS state (a) and the
two-mode squeezed vacuum state (b). 
Parameters for (a) are:
reflectance $r^2_{m=0}=0.5$, $d=100$ km, 
central wavelength $\lambda_c=1560$ nm,
repetition frequency $\omega_{rep}/2\pi=250$ MHz, 
pulse width $0.5$ ns, 
number of comb teeth $2M+1=11$,
central comb tooth $m=0$, 
spectral weight $\eta_{m=0}=0.2$, 
diffraction-limited beam divergence $\theta=1 \mu$rad, 
Rayleigh range $z_0=180$ km,
atmospheric roundtrip attenuation $\ell_{m=0}=0.64$, 
overlaps $O_{{\cal S}}=O_{{\cal I}}=1$,
collection efficiency $\mu_{coll} = 65 \%$, and
detector efficiency $\mu_d=90 \%$. 
Squeezing amplitude and phase are $g=2.3 \equiv 20$ dB and $\phi_g=\pi$. Probe signal photon propagation phase is
$\varphi_{s_r}=\pi/2$. Other phases are $\varphi_i=\varphi_{S}=\varphi_I= \phi_r=\phi_{\xi}=0$.
In (b), $g=2.3$ and $\phi_g=\pi$. Note the different scales for $x_+$ and $x_-$ in (b).
}
\label{Fig2}
\end{figure}

Although the two TMSVc used in qCOMBPASS are Gaussian states, 
the qCOMBPASS transceiver state Eqs. (\ref{transceiver}) is non-Gaussian
because idler and signal pulses of the forward TMSVc are subtracted by the measurement and by the chopper, respectively  \cite{Ra2020,Walschaers2021}. 
Fig. \ref{Fig2} shows the Wigner function of the qCOMBPASS transceiver state
in the $(x_+, x_-)$ quadrature space. The inset highlights the squeezed nature of the state (squeezing along $x_+$ for the parameters in the figure) and the displacement from the origin.

The Wigner function of the qCOMBPASS transceiver state is calculated as
\cite{Agarwal}

%%%%%%%%%%%%%%%
%%%%%%%%%%%%%%%

%
\begin{eqnarray}
W_{\rm qCOMBPASS}(\alpha,\beta) &=& \frac{4}{\pi^4} \; e^{2 (|\alpha|^2+|\beta|^2)} 
 \\
&\times&
\int d^2u \int d^2v \;
e^{2 (u^* \alpha + v^* \beta - u \alpha^* - v \beta^*)}
\langle - u| _{\cal S}  \langle -v|_{\cal I} \;
\rho_{{\rm qCOMBPASS}} \; 
|v\rangle_{\cal I}
| u\rangle _{\cal S} 
\nonumber 
\end{eqnarray}
where $\alpha$ and $\beta$ are complex numbers and
$|u \rangle$ and $|v \rangle$ are coherent states. Expanding the coherent states in the Fock basis and using Eq. (\ref{rhocombpass}) we find
\begin{eqnarray}
&& W_{\rm qCOMBPASS}(\alpha,\beta) = \frac{4}{\pi^4} \;  e^{2 (|\alpha|^2+|\beta|^2)}
(1-x^2)^2 (1-y^2) \nonumber \\
&& \times 
\sum_{n,n'} \sum_{p,p'} \sum_{q,q'} 
\frac{
e^{i [ (n-n') (\varphi_S + \varphi_I - \phi_g) + (p-p') (\varphi_i + \phi_g) 
+ (q-q') (\varphi_{s_r} + \phi_r + \phi_g + \phi_{\xi})] }
x^{n+p+n'+p'} \; y^{q+q'} (-1)^{p+q}
}
{\sqrt{(n+q)! \; (n'+q')! \; (n+p)! \; (n'+p')
! } }
\nonumber  \\
&& \times
J(\alpha,\beta;n+q, n'+q',n+p,n'+p') 
\end{eqnarray}
where
\begin{eqnarray}
&& J(\alpha,\beta;n+q, n'+q',n+p,n'+p') \nonumber \\
&&
= \frac{1}{\pi^2}
\int d^2u \int d^2v \;
e^{- |u|^2 - |v|^2} \; 
e^{2 (u^* \alpha + v^* \beta - u \alpha^* - v \beta^*)}
(u^*)^{n+q} u^{n'+q'}  (v^*)^{n+p} v^{n'+p'}  
\nonumber \\
&&
=\frac{(-1)^{p'+q'}}{2^{2n+2n'+p+q+p'+q'}}
\left(  \frac{\partial}{\partial \alpha}\right)^{n+q} \!
\left(  \frac{\partial}{\partial \alpha^*}\right)^{n'+q'} \!
\left(  \frac{\partial}{\partial \beta}\right)^{n+p} \!
\left(  \frac{\partial}{\partial \beta^*}\right)^{n'+p'}
e^{-4 ( |\alpha|^2 + |\beta|^2 )}
\end{eqnarray}
Introducing quadrature operators 
\begin{eqnarray}
\hat{x}_{\cal S} = \frac{1}{\sqrt{2}} (a_{\cal S} + a^{\dagger}_{\cal S}) &; &
\hat{y}_{\cal S}  = \frac{1}{i\sqrt{2}} (a_{\cal S} - a^{\dagger}_{\cal S}) \nonumber \\
\hat{x}_{\cal I}  =\frac{1}{\sqrt{2}} (a_{\cal I} + a^{\dagger}_{\cal I}) &; &
\hat{y}_{\cal I}  =  \frac{1}{i \sqrt{2}} (a_{\cal I} - a^{\dagger}_{\cal I})
\end{eqnarray}
with corresponding eigenvalues $x_{\cal S}$,  $y_{\cal S}$,  $x_{\cal I}$,  $y_{\cal I}$,
and coordinates $x_{\pm} = (x_{\cal S} \pm x_{\cal I})/\sqrt{2}$ and
$y_{\pm} = (y_{\cal S} \pm y_{\cal I})/\sqrt{2}$, we finally get an expression for the qCOMBPASS Wigner function
$W(x_+, x_-, y_+, y_-)$. In Fig. 2 we compare the Wigner function for the non-Gaussian two-mode qCOMBPASS transceiver state with the squared Wigner function of the Gaussian two-mode squeezed vacuum (TMSV) state. The latter is given by the simple formula \cite{Agarwal}
\begin{eqnarray}
W_{TMSV}(\alpha,\beta) = \frac{4}{\pi^4} \;
e^{ -2 | \alpha \cosh(g) - \beta^* \sinh(g) e^{i \phi_g}|^2 } \;
e^{ -2 | -\alpha^* \sinh(g) e^{i \phi_g} - \beta \cosh(g)|^2 }
\end{eqnarray}
This is Gaussian state centered at the origin and squeezed along the $x_+$ quadrature for the parameters in Fig. 2(b). For qCOMBPASS, the Wigner function is  displaced from the origin and is also squeezed along
the $x_+$ quadrature, although squeezing is less pronounced than for the TMSV state Fig. 2(a).

%%%%%%%%%%%%%

\vspace{0.2cm}

\section
{Distance dependence of the reflected signal field}

Suppose the signal field at $z=0$ is a linearly polarized Gaussian beam with electric field amplitude
\begin{equation}
E(x,y,0)=E_0 \; e^{-(x^2+y^2)/w_0^2}
\end{equation}
Assuming a Gaussian probe pulse at the exit of the qCOMBPASS station with carrier wavelength $\lambda_s$ and beam waist $w_0$,
the Rayleigh range for free-space propagation is $z_0=\pi w_0^2/\lambda_{m=0}$ and
the transverse extent of the field is $w(z)=w_0 (1+ z^2/z_0^2)^{1/2}$. 
Each comb tooth suffers slightly different diffraction according to its wavelength,
$\lambda_m= 2\pi c/\omega_{s,m}$.
For a target at $z=d$ and zero zenith angle, the field after reflection from the target and propagation back to $z=0$ is
\begin{equation}
E_{back}(x,y,0)=r e^{i \phi_r} E_0\frac{w_0}{w(2d)}e^{-(x^2+y^2)/w^2(2d)}e^{ik(x^2+y^2)/2R(2d)}\tan^{-1} \left( \frac{2d}{z_0} \right)
\label{fraun1}
\end{equation}
%^
where $r e^{i \phi_r}$ is the target's complex reflection coefficient, $k=2\pi/\lambda$, $z_0=\pi w_0^2/\lambda$, $w(2d)=w_0(1+4d^2/z_0^2)^{1/2}$, and $R(2d)=2d+z_0^2/2d$. For $d=100$ km, $\lambda=1560$ nm, and $w_0=30$ cm, for example, $w \cong 2\lambda d/\pi w_0$, $R \cong 2d$, and
\begin{equation}
E_{back}(x,y,0)\cong \frac{r E_0 \lambda}{2\pi \theta^2 d} 
e^{-\frac{x^2+y^2}{4 \pi \theta^2 d^2}} 
e^{ik(x^2+y^2)/2d}
\end{equation}
where $\theta=\lambda/\pi w_0$ is the beam divergence.
For the numbers assumed, $\pi w_0^2/2\lambda d \approx 0.9$ and the spot size at the target is $2\lambda d/\sqrt{\pi}w_0 \approx 0.6$ m. The strength of the return signal field can obviously be significantly increased by increasing $w_0$ and decreasing the wavelength $\lambda$. The spot size at the target must be small compared to the $x,y$ dimensions of the target in order to justify our simplifying assumption that the target intercepts the entire field incident upon it. For the numbers assumed, this requires that the $x,y$ dimensions of the target be larger than a few meters. More generally, the returning signal field is determined by a Fraunhofer integral over the target's $xy$ area. This results in a diminished return signal compared with the ``large target" scenario.

The illumination spot at the target position is $w(d)$. On the return, 
we take as new initial beam waist $w'_0 \approx w(d)$, the new Rayleigh range is
$z'_0=\pi w_(d)^2/\lambda_{m=0}$, and the reflected illumination
spot at the qCOMBPASS base position is 
$w'(0)=w'_0 (1+ d^2/z_0^{'2})^{1/2}$.
Assuming full illumination of the target (target geometric cross-section larger than $w(d)$)
the renormalized squeezing parameter is approximated as
$\tilde{g}=g \mu_{coll} r \ell z_0/d$. 
For parameters
$g=1.7$ (15 dB squeezing), 
$\eta_{m=0}=0.2$, 
$\lambda_{m=0}=1560$ nm,
$r^2=0.5$ (Si target), 
$\ell \approx 0.8\times 0.8=0.64$ (roundtrip atmospheric loss),
$w_0=30$ cm (achieved via a telescopic beam expander), and
$d=100$ km, 
we get
$z_0 \approx 180 \; {\rm km}$,
$w(d) \approx  34\; {\rm cm}$, 
$z'_0 \approx 230 \; {\rm km}$,
and $w'(0) \approx  37\; {\rm cm}$.
The collection efficiency is $\mu_{coll}=[w_0/w'(0)]^2 \approx 0.65$. Hence
$\tilde{g} \eta_{m=0}\approx 0.2$.

The $1/d$ dependence of the peak amplitude of $E_{back}(x,y,0)$ is a consequence of the interfering spherical Huygens wavelets from different points $(x,y)$, corresponding to the factor $\exp{[ik(x^2+y^2)/2R]}$ in the integration over $x$ and $y$ in the Fraunhofer diffraction formula. This results in a factor $d$ that multiplies a factor $1/d^2$ in the Fraunhofer formula for $E_{back}(x,y,0)$. This is in contrast to the $1/d^2$ dependence of the return signal in LIDAR, which follows from an $xy$ integration over the field intensity under the assumption that each point $x,y$ at the target acts as an independent point source proportional to the intensity. This gives a return LIDAR signal proportional $P/d^2$, where $P$ is the field power.

%%%%%%%

\vspace{0.2cm}

\section
{Effects of atmospheric turbulence on ground-based qCOMBPASS}

Atmospheric turbulence, which we have thus far ignored in our analyses and numerical estimates, will of course affect the propagation of the signal field. Turbulence causes the beam to bend and possibly miss the target. It also causes beam spreading beyond the diffraction limit. We can make rough estimates of the magnitude of these two effects as follows. For present purposes we consider only ``clear air" turbulence and ignore, among other things, effects of clouds, fog, and aerosols. 

From the Kolmogoroff theory of isotropic and homogeneous turbulence it may be assumed for our purposes  that the variance in the refractive index fluctuations between two points is $C_n^2r^{2/3}$, where $r$ is the separation between the two points and $C_n^2$ is the refractive index structure constant \cite{Goodman1985}. $C_n^2$ varies widely depending on atmospheric conditions, altitude, and other factors, and is typically of order $10^{-17}$ m$^{-2/3}$ and $10^{-13}$ m$^{-2/3}$ for weak and strong atmospheric turbulence, respectively, near ground level. The effects of turbulence on beam propagation may be characterized by the coherence diameter \cite{fried}
\begin{equation}
r_0=0.186 \left[ \frac{\lambda^2}{\int_{d_i}^{d}dzC_n^2(z)} \right]^{3/5}
\end{equation}
where $d-d_i$ is the total propagation length. $r_0$ is typically around 20 cm and 5 cm for weak and strong turbulence, respectively \cite{Goodman1985}.
The rms lateral beam displacement due to turbulence has been found to be approximately \cite{fante,weichel,andrews}
\begin{equation}
\sqrt{\langle\rho_w^2\rangle}\approx \frac{\lambda d}{2w_0} \left(\frac{w_0}{r_0} \right)^{5/6}
\end{equation}
which is on the order of a meter for $r_0=5$ cm, $\lambda=1560$ nm, and 
$w_0=30$ cm. Thus, beam wander should not be a serious issue for target dimensions and receiver apertures of a few meters or larger.
The rms beam spread $\sqrt{\langle\rho_s^2\rangle}$ can be roughly estimated as $2\lambda d/\pi r_0$ in a ``worst case" in which $r_0\ll w_0$. Thus, for $\lambda=1560$ nm, $d=100$ km, and $r_0=5$ cm,  $\sqrt{\langle\rho_s^2\rangle}\approx 1.2$ m, suggesting that beam spreading due to turbulence is a small effect for target dimensions greater than a few meters.

%%%%
\vspace{0.2cm}

\section{The qCOMBPASS equation} 

The mean number of idler photons 
$N_{{\cal I},m}={\rm Tr}[ a^{\dagger}_{{\cal I},m} a_{{\cal I},m} \rho_{{\rm qCOMBPASS}}]$
measured by the idler detector at comb frequency $\omega_{i,m}$ at times when the chopper is on is 
given by the {\it qCOMBPASS equation} 
\\
\begin{eqnarray}
N_{{\cal I},m}&=&  
2 \mu_d  \sinh^2(g \eta_m) 
\left\{
1+
\frac{2 O_{{\cal S}} O_{{\cal I}}  \tanh(\tilde{g}_m \eta_m) [\cos \Phi_m- \tanh(\tilde{g}_m \eta_m)]}
{1+\tanh^2(\tilde{g}_m \eta_m)-2 \tanh(\tilde{g}_m \eta_m) \cos\Phi_m}
\right\}
\label{qCOMBPASS}
\end{eqnarray}
Here, $\mu_d$ is the detection efficiency of reflected signal photons and the phase 
$\Phi_m=\varphi_{s,m}(d)+ \varphi_{i,m}- \varphi_{S,m}-\varphi_{I,m}+\phi_{r,m}+ \phi_{\xi,m}+\phi_g$ contains propagation, reflection, noise, and quadrature squeezing  phases, all evaluated at the corresponding signal or idler comb frequencies. Assuming 
coherent full illumination, i.e., a target geometric cross section larger than the illumination 
spot size at the target,
the renormalized squeezing amplitude is 
$\tilde{g}_m=g \mu_{coll} r_m \ell_m z_{0,m}/d$.
The $1/d$ fall-off with distance-to-target arises from the product of a $1/d$ electric field decay factor 
in the trip to target, a $d$ factor due to coherent full illumination of the target, and
another $1/d$ factor due to re-irradiation (reflection) for the return trip.
Assuming full illumination of the target, the renormalized squeezing amplitude is
$\tilde{g}_m=g \mu_{coll} r_m \ell_m z_{0,m}/d$, where
$\mu_{coll}$ is the collection efficiency of reflected signal photons,
$r_m$ is the target reflectivity, and
$\ell_m =\exp[- 2 \int_0^d \alpha_{atm}(z,\omega_{s,m}) dz]$ is the atmospheric attenuation.
The Rayleigh range for  a comb tooth signal pulse 
of Gaussian transverse spatial profile is $z_{0,m}=\pi w^2_{0,m} /\lambda_m$,
where the initial beam waist at exit from the qCOMBPASS station is
$w_{0,m}$ and $\lambda_m=2\pi c/\omega_{s,m}$ is the comb tooth wavelength.
In perturbative qCOMBPASS $g \ll 1$ and in the near-field regime 
(no far-field fall-off, no diffraction), Eq.\,(\ref{qCOMBPASS}) reduces to
Eq.\,({\ref{spdc}). 
For strong non-perturbative qCOMBPASS $g \eta_m, \tilde{g}_m \eta_m \gg 1$ and in the far-field regime,
the photo-counts and visibility are
$N_{{\cal I},m} \sim e^{2g \eta_m}$ and $V_{{\cal I},m}\approx1-2 e^{-4\tilde{g}_m\eta_m}$
for $O_{{\cal S}}= O_{{\cal I}}=1$. 
See Appendix for a detailed derivation of the qCOMBPASS equation.

%%%%%%%%%%%%%%%%
\begin{figure}[t]
\includegraphics[width=0.8 \linewidth]{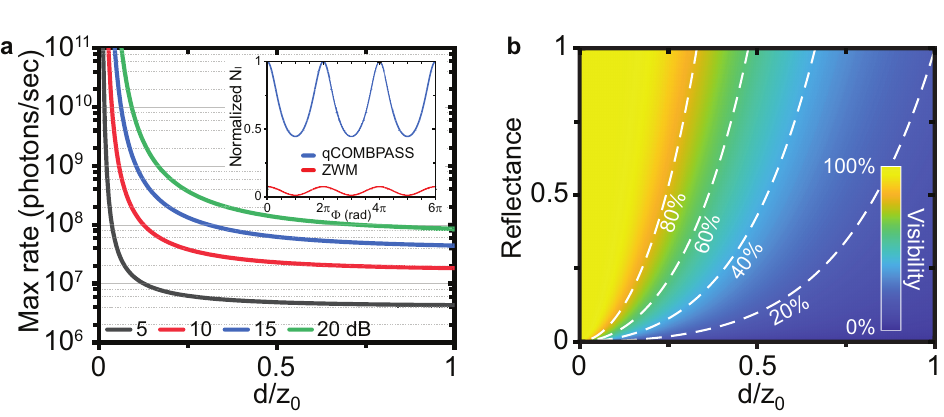}
\caption{
{\bf qCOMBPASS equation:}
(a) qCOMBPASS maximum photo-counts vs distance-to-target over diffraction Rayleigh range $d/z_0$ for various squeezing levels. 
Inset: Normalized interference patterns for remote-range 
non-perturbative qCOMBPASS ($d=100$ km, $g=1.7 \equiv 15$ dB) and 
near-range cw perturbative ZWM ($g=0.1\equiv 0.9$ dB). Normalization is  
the maximum of qCOMBPASS photo-counts. 
Photon loss and diffraction are assumed negligible in near-range sensing; 
(b) qCOMBPASS visibility vs target reflectivity and distance-to-target for $g= 1.7$;
(d) Remote quantum metrology: Region of parameter space $(\Phi,g)$
where $\Delta \Phi_{\rm qCOMBPASS} / \Delta \Phi_{SQL} < 1$.
Inset: Phase uncertainty at operating point $\Phi=\pi/4$. 
Unless quoted otherwise, parameters in the various panels are:
central wavelength $\lambda_c=1560$ nm,
repetition frequency $\omega_{rep}/2\pi=250$ MHz, 
pulse width $0.5$ ns, 
number of comb teeth $2M+1=11$,
central comb tooth $m=0$, 
spectral weight $\eta_{m=0}=0.2$, 
diffraction-limited beam divergence $\theta=1 \mu$rad, 
Rayleigh range $z_0=180$ km;
target reflectance $r^2_{m=0}=0.5$, 
distance $d=100$ km,
atmospheric roundtrip attenuation $\ell_{m=0}=0.64$, 
overlaps $O_{{\cal S}}=O_{{\cal I}}=1$,
collection efficiency $\mu_{coll} = 65 \%$, and
detector efficiency $\mu_d=90 \%$. 
}
\label{Fig3}
\end{figure}
%%%%%%%%%%%%%%%

Details of the derivation of the qCOMBPASS equation can be found in the Appendix.
This  equation is a quantum analog of the well-known LIDAR  equation that describes detection and ranging using classical light. Importantly for remote sensing, the decay with distance is milder than the  $1/d^2$  fall-off in LIDAR.
Note that when the chopper is off and no reflected signal photons are arriving at the qCOMBPASS station, the idler detector still measures a photo-count baseline 
$N^{(baseline)}_{{\cal I},m}= 2 \mu_d  \sinh^2(g \eta_m)$ due to idler photons $i, I$ that do not contain target information. 
Fig. 3a shows the maximum number of photo-counts as a function of distance-to-target for increasing squeezing, showing how non-perturbative qCOMBPASS mitigates the fall-off of the signal with distance. The inset 
compares the interference patterns of non-perturbative qCOMBPASS and perturbative ZWM, showing that the former has a much larger visibility.
The visibility of the qCOMBPASS interference pattern is
\begin{eqnarray}
V_{{\cal I},m}&=&
\frac{2 O_{{\cal S}} O_{{\cal I}}   \tanh(\tilde{g}_m \eta_m) }{ 1+ O_{{\cal S}} O_{{\cal I}}  \tanh^2(\tilde{g}_m \eta_m)}
\label{qCOMBPASSvisibility}
\end{eqnarray}
Fig. 3b depicts the qCOMBPASS visibility as a function of the reflectance and 
distance-to-target for fixed squeezing $g=1.7$ ($15$ dB). Low-reflectivity remote targets can be sensed using highly-squeezed probes. 
Fig. 4 shows visibility versus distance and reflectance.
For strong squeezing $g \eta_m, \tilde{g}_m \eta_{m} \gg 1$
(e.g., highly-collimated, high-power signal beams with small atmospheric attenuation),
the qCOMBPASS equation predicts an exponentially large photo-count  
$N_m \sim e^{2g \eta_m}$ and an almost unit visibility. For large unrenormalized squeezing $g \eta_m \gg 1$ but small renormalized squeezing 
$ \tilde{g}_m \eta_{m} \ll 1$ (e.g., high-power but strong dissipation),
$N_m \sim e^{2 g \eta_m}$ but the visibility is small $V_m = 2 \tilde{g}_m \eta_m$. Finally, when both squeezing amplitudes are small (e.g., low-power pump), we recover the perturbative solution of Eqs. (\ref{spdc}).
For the present purpose of introducing the qCOMBPASS equation, we assumed for simplicity that pump depletion in the nonlinear crystal can be ignored. However, at strong pump power, depletion can saturate the number of emitted photons \cite{depletion} and limit squeezing levels, currently 
\cite{Vahlbruch2016} at $\lesssim 15$ dB, which corresponds to $g \lesssim 1.7$.

%%%%%%%%%%%%%%%%
\begin{figure}[t]
\includegraphics[width=0.8 \linewidth]{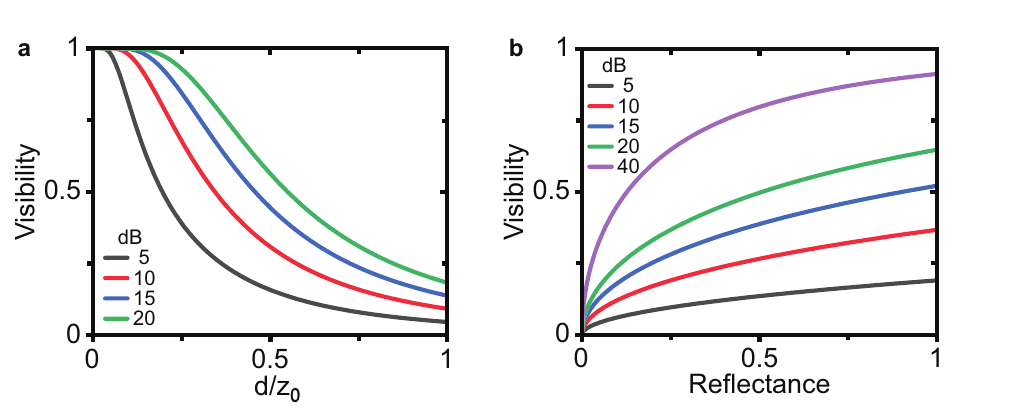}
\caption{{\bf qCOMBPASS visibility.}
Visibility as a function of distance (a) and of reflectance (b). 
In (a) the reflectance is $r^2_{m=0}=0.5$ and in (b) the distance is $d=100$ km. Other parameters are:
central wavelength $\lambda_c=1560$ nm,
repetition frequency $\omega_{rep}/2\pi=250$ MHz, 
pulse width $0.5$ ns, 
number of comb teeth $2M+1=11$,
central comb tooth $m=0$, 
spectral weight $\eta_{m=0}=0.2$, 
diffraction-limited beam divergence $\theta=1 \mu$rad, 
Rayleigh range $z_0=180$ km,
atmospheric roundtrip attenuation $\ell_{m=0}=0.64$, 
overlaps $O_{{\cal S}}=O_{{\cal I}}=1$,
collection efficiency $\mu_{coll} = 65 \%$, and
detector efficiency $\mu_d=90 \%$. }
\label{Fig4}
\end{figure}
%%%%%%%%%%%%%%%

Hyperspectral remote quantum sensing and imaging is also possible with qCOMBPASS
because 
detection at an idler comb frequency $\omega_{i,m}$ contains target information
at the corresponding signal comb frequency $\omega_{s,m}$.

%%%
\vspace{0.2cm}

\section{Remote quantum metrology with qCOMBPASS} 

The uncertainty in phase estimation depends on the variance and  derivative 
of the photo-counts as \cite{Giovannetti2011}
\begin{equation}
\Delta \Phi_{{\rm qCOMBPASS},m} =\frac{\sqrt{ \langle N^2_{{\cal I},m} \rangle -  \langle  N_{{\cal I}.m} \rangle^2}}
{\left| \langle  \frac{\partial N_{{\cal I},m}}{\partial \Phi} \rangle \right|},
\end{equation} 
where $N_{{\cal I}.m}$ is given by the qCOMBPASS equation Eq. (\ref{qCOMBPASS}), see Appendix. 
The expectation values are given by
\begin{eqnarray}
\langle N_{{\cal I},m} \rangle^2 &=&
\frac{4x_m^4}{(1-x_m^2)^2} 
\left[1+\frac{2y_m (\cos \Phi- y_m)}{1+y_m^2-2y_m \cos\Phi} \right]^2
\nonumber \\
\left| \langle \frac{\partial N_{{\cal I},m}}{\partial \Phi} \rangle \right|
&=&
\frac{4 x_m^2 y_m (1-y_m^2) }{(1-x_m^2) (1+y_m^2-2y_m \cos\Phi)^2} \; |\sin\Phi |
\nonumber \\
\langle N^2_{{\cal I},m} \rangle  &=&
\frac{2x_m^2 (1+2x_m^2)}{(1-x_m^2)^2}  
\left[
1+ \frac{2y_m (\cos \Phi-y_m)}{1+y_m^2-2y_m \cos\Phi}
\right] 
= \frac{1+2 x_m^2}{ 1-x_m^2} \; \langle N_{{\cal I},m} \rangle
\end{eqnarray}
where $x_m=\tanh( g\eta_m)$ and  $y_m=\tanh( \tilde{g}\eta_m)$.
In Fig. 4(a) we show the region of parameter space where the ratio 
$\Delta \Phi_{{\rm qCOMBPASS},m=0}  / \Delta \Phi_{SQL,m=0}$ becomes smaller than 1
and the inset corresponds to an operating point $\Phi=\pi/4$ where the qCOMPASS is more precise than SQL in the range $1 \lesssim g \lesssim 4$ for the parameters in the figure. Although here we have described phase detection via  photo-counting, other schemes based on homodyne or heterodyne detection of idler photons are possible to measure phase shifts with precision beyond SQL.
In this regime of parameters 
qCOMBPASS can estimate phases with sensitivity and resolution better than a classical probe, 
There is an optimal operating point $(\Phi,g)_{opt}$ where the difference 
$\Delta \Phi_{SQL,m=0} - \Delta \Phi_{{\rm qCOMBPASS},m=0} \approx 2$
is maximal.

%%%%%%%%%%%%%%%%
\begin{figure}[t]
\includegraphics[width=1 \linewidth]{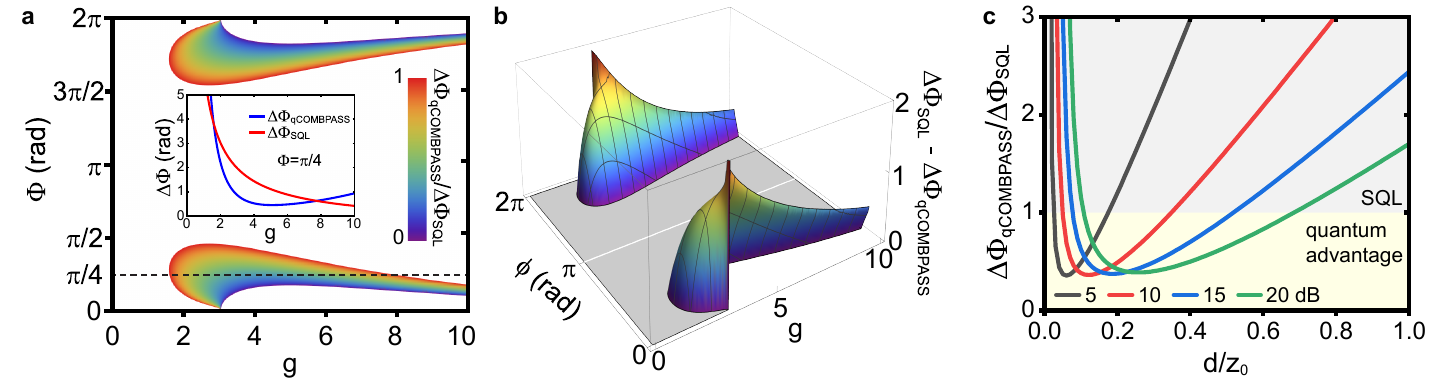}
\caption{
{\bf Remote quantum metrology:}
(a)  Region of parameter space $(\Phi,g)$
where $\Delta \Phi_{{\rm qCOMBPASS},m} / \Delta \Phi_{{SQL},m} < 1$.
Inset: Phase uncertainty at operating point $\Phi=\pi/4$.
(b) Region in the $(\Phi,g)$ plane where  $\Delta \Phi_{{SQL},m} - \Delta \Phi_{{\rm qCOMBPASS},m}>0$.
Parameters are:
central wavelength $\lambda_c=1560$ nm,
repetition frequency $\omega_{rep}/2\pi=250$ MHz, 
pulse width $0.5$ ns, 
number of comb teeth $2M+1=11$,
central comb tooth $m=0$, 
spectral weight $\eta_{m=0}=0.2$, 
diffraction-limited beam divergence $\theta=1 \mu$rad, 
Rayleigh range $z_0=180$ km;
target reflectance $r^2_{m=0}=0.5$, 
distance $d=100$ km,
atmospheric roundtrip attenuation $\ell_{m=0}=0.64$, 
overlaps $O_{{\cal S}}=O_{{\cal I}}=1$,
collection efficiency $\mu_{coll} = 65 \%$, and
detector efficiency $\mu_d=90 \%$. 
(c) Quantum advantage as a function of distance for different squeezing levels.
All parameters are the same as in (a). Operating point is $\Phi=\pi/4$.
}
\label{Fig5}
\end{figure}
%%%%%%%%%%%%%%%

Furthermore, squeezing causes the nonlinear dependency of the qCOMBPASS equation 
in the interference factor $\cos\Phi_m$. As a result, average over noise realizations does not vanish  
Eq.\,(\ref{qCOMBPASS}), giving robustness again random noise.
qCOMBPASS can be used to measure phases
with precision better than is classically possible and reach uncertainties below the standard quantum limit (SQL), also known as the shot-noise limit  \cite{Giovannetti2011}. For example, it can measure a small displacement of the target, reflected in a small shift of the signal photons' propagation phase, with resolution beyond the SQL.
The qCOMBPASS phase uncertainty is determined as 
$\Delta \Phi_{{\rm qCOMBPASS},m} =\sqrt{ \langle N^2_{{\cal I},m} \rangle -  \langle  N_{{\cal I},m} \rangle^2} / \left|  \langle \frac{\partial N_{{\cal I},m}}{\partial \Phi}  \rangle \right|$.
In a certain region of parameter space $(\Phi,g)$, qCOMBPASS performs better than SQL, i.e., 
\begin{equation}
\Delta \Phi_{{\rm qCOMBPASS},m}  < \Delta \Phi_{SQL,m} 
\end{equation}
where $\Delta \Phi_{SQL,m}=1/\sqrt{\overline{N}_{s,m}}$ and
$\overline{N}_{s,m}=\sinh^2(\tilde{g}_m \eta_m)$ is the mean number of photons in the signal pulse at comb frequency $\omega_{s,m}$.
As a consequence, qCOMBPASS provides quantum advantage for remote quantum metrology.

%%%%%%%%%%%%%%
\vspace{0.2cm}
\section{qCOMBPASS Experimental schemes}

Perturbative qCOMBPASS can be implemented using a classical frequency comb in the low-power regime, as shown in Fig. 6a, that corresponds to an actual realization of the simple model of Fig. 1. In a forward pass,
a frequency comb pumps a single PPLN crystal and produces collinear, non-degenerate signal-idler pulses of the same linear polarization (type-0 SPDC). This setup is basically one-dimensional, making alignment for path identity simpler without the need of Y-junctions. After a forward pass, pump, signal, and idler pulses are spatially separated by dichroic mirrors and
pump and idler pulses are retro-reflected. Trains of $p$ consecutive signal pulses are blocked or unblocked by the chopper
before propagating to the  target.
In a reverse pass, pump pulses generate new signal-idler pairs that are aligned with retro-reflected idler pulses and with signal pulses 
reflected off the target. Finally, reverse pump, signal, and idler pulses are separated with dichroic mirrors, and an idler detector measures idler pulses. Signal pulses are not measured but are absorbed in a beam blocker after idlers are detected (signal photons could also be measured). 
Target information can be retrieved from idler pulses via ZWM interference, although they never went to the target or interacted with probe signal photons that did go to the  target. Moving the idler retro-reflecting mirror enables the buildup of the qCOMBPASS interference pattern when the chopper is ON (Eq. (\ref{spdc})). Note there is no quantum memory in the setup to store any single photon. In this low-power SPDC regime, the generated biphoton frequency combs contain only one photon per pulse with small probability amplitude $g$. As such, this scheme is not suitable for quantum remote sensing because the probe photon can easily be lost in the atmosphere.
Perturbative qCOMBPASS is only suitable for near-range sensing and imaging. 

Pulsed lasers with no  phase stabilization might provide a poor-man's version of frequency combs for limited time-spans and keep 
some level of pulse-to-pulse coherence for a simple qCOMBPASS demonstration. Frequency combs with carrier-envelope offset (CEO) frequency stabilized to a non-zero value could be used for qCOMBPASS as long as intra- and inter-comb phase-locking is preserved between old and new pulses for long time spans. However, the best qCOMBPASS performance is expected for frequency combs with zero carrier-envelope offset in which all pulses are nominally identical.

Our team has started an experimental program to demonstrate qCOMBPASS. To this end, we first conducted a quantum imaging tabletop experiment on ZWM in the perturbative regime using a continuous wave (cw) laser pump. The results are reported in Fig. 6b. Conceptually, the setup is 
similar to Fig. 6a with the  difference that the pumping laser is cw instead of a frequency comb, the retro-reflecting mirror for the pump is not movable, and we use a different nonlinear crystal. 
We employ a 405 nm pump laser, power 12 mW, beam diameter 500 $\mu$m, and the poling period of the PPKTP type-0 crystal is 3.425 $\mu$m. The wavelengths of SPDC photon pairs are 783 nm and 840 nm. We use a EMCCD camera (Teledyne ProEM-HS:1024BX3) to detect low-photon flux of idler photons. In experiment we scan the movable mirror by 80 ${\mu}$m with 200 nm steps and a 30 s exposure time. The visibility is computed as 
$V=\sqrt{2 \sum_{n=1}^N |{\cal F}(q_n)|^2}/{\cal F}(q_0=0)$, where ${\cal F}(q_n)$ is the spatial Fourier transform of the photocount as a function of mirror position, $q$ is momentum, $n$ sums over discrete momenta, and $N$ is a cutoff momentum.
We emphasize that the experiment reported in Fig. 6b is {\it not} a demonstration of qCOMBPASS (that requires frequency combs), but simply an experiment with a CW pump showing how the well-known ZWM phenomenon works. 
The goal of this experiment is for our experimental team to gain expertise in quantum-induced coherence by path identity, and we decided to report our first results for the ZWM approach in this paper. 

When the phase matching conditions of the crystal can be satisfied over a broad frequency range,
the emitted signal/idler pulses do not form frequency combs, but are a superposition of various pairs of frequencies that do not have the comb dispersion $\omega_{s/i,m}=\omega_c+ m \omega_{rep} + \omega_{ceo}$. Here, $\omega_c$ is the center frequency, $\omega_{rep}$ is the repetition frequency, and 
$\omega_{ceo}$ is the carrier-envelope-offset
frequency.
To remedy this problem, reference \cite{Lee2018} introduced an ultra-narrow cw seedling laser at a frequency verifying the matching conditions. This makes the crystal preferentially emit idler photons at a fix frequency, while the companion signal photons are still a quantum frequency comb.

%%%%%%%%%%%%%%%
\begin{figure}[t]
\includegraphics[width=0.8\linewidth]{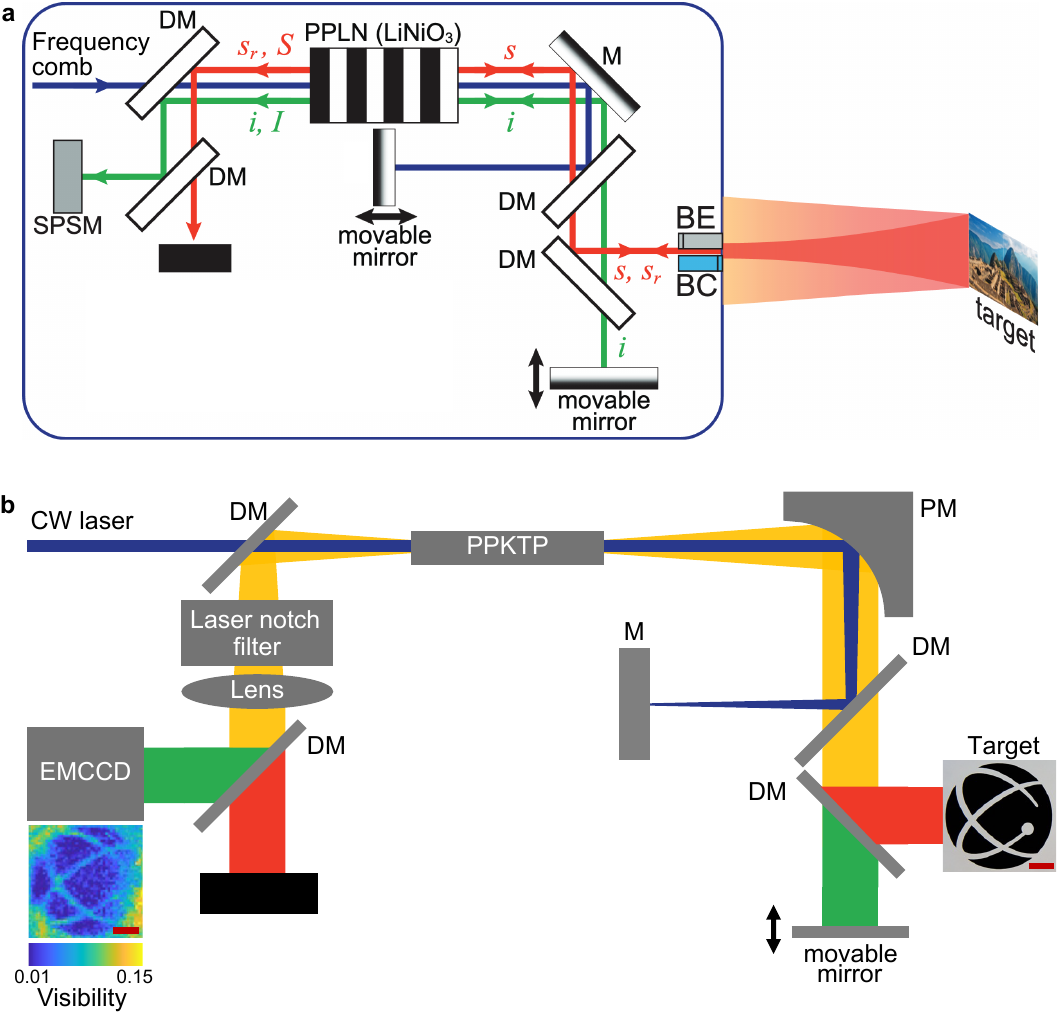}
\caption{{\bf  (a) Schematics of perturbative qCOMBPASS experimental setup.}
A classical frequency comb pumps a periodically poled lithium niobate crystal (PPLN) producing collinear non-degenerate photon pairs via type-I SPDC. 
BC (beam collector), 
BE (beam expander), DM (dichroic mirror), 
M (mirror),
SPSM (single photon spectrometer).
{\bf (b) Perturbative ZWM experimental results using a cw pump laser.}
Retrieved image (left) from idler photons that never interact with target (right, 100 nm thick Cr mask with
Los Alamos National Laboratory logo).
Scale bars of target and image are 500 $\mu$m. Distance to target is approximately 3 cm.
EMCCD (electron-multiplying CCD), PM (parabolic mirror), PPKTP crystal (potassium titanyl phosphate).
}
\label{Fig6}
\end{figure}
%%%%%%%%%%%%%%%

In Fig. 7 we show a possible experimental setup to demonstrate non-perturbative qCOMBPASS.
A frequency comb pump with center frequency $\omega_c$, repetition frequency $\omega_{rep}$, and zero carrier-envelope-offset
frequency $\omega_{ceo}=0$,  undergoes second-harmonic generation (SHG) and is then split into four beams to pump four identical 
PPKTP degenerate optical parametric amplifiers (DOPAs). The free spectral range
of the bow-tie DOPA resonators is matched to the input repetition rate as $\omega_{rep}=2\omega_{FSR}$, enabling the generation of
identical single-mode squeezed vacuum combs (SMSVc) in each DOPA with degenerate signal/idler comb frequencies 
$\omega_{s,m}=\omega_{i,m}=\omega_c/2 + m \omega_{FSR}$. 
The cavity length is held on resonance for the fundamental and pump fields via a 
high-bandwidth Pound-Drever-Hall (PDH)
locking scheme (not shown). Pairs of DOPAs are used to produce two-mode squeezed vacuum combs by
 phase shifting by $\pi/2$ one of the SMSVc's before mixing in a 50-50 beam splitter. The two pairs of DOPAs form
the  ``forward"  and  ``reverse" TMSVc with modes $(s,i)$ and $(S,I)$, respectively. 
An alternative to using a pair of DOPAs to produce a TMSVc is to employ a single PPKTP crystal inside a Sagnac interferometer pumped in both clockwise and counter-clockwise directions simultaneously \cite{Pan2020} and whose two output ports are two degenerate SMSVc combs that are later combined in the beam splitter to form the TMSVc.
A mechanical chopper blocks/unblocks trains of  $p$ consecutive signal probe pulses from the forward TMSVc to implement the chopper protocol described above. A beam expander is used to extend the beam transverse size before exiting the qCOMBPASS station to minimize diffraction, and a high NA lens beam collector captures  light reflected off the target.
Delay lines are used to ensure temporal overlap between $s_r$ and $S$ pulses before aligning them in a fiber-based Y-junction for path identity, and idler pulses $i$ and $I$ are temporally overlapped and mixed in another Y-junction. 
A single-photon spectrometer is used for
frequency-resolved detection of idler photons at comb frequencies $\omega_{i,m}$  to obtain hyperspectral target information via quantum-induced coherence by path identity. 
The delay line for idler pulses $i$ is used to build the qCOMBPASS interference pattern. 
Small horizontal shifts of the pattern due to possible target displacements provide differential time-of-flight information. Furthermore, Doppler-shifts of the reflected comb lines give a precise tool for 
target vectoring, e.g., determining linear and rotational velocities of an object of interest.

%%%%%%%%%%%%%%%
\begin{figure}[t]
\includegraphics[width=1.0\linewidth]{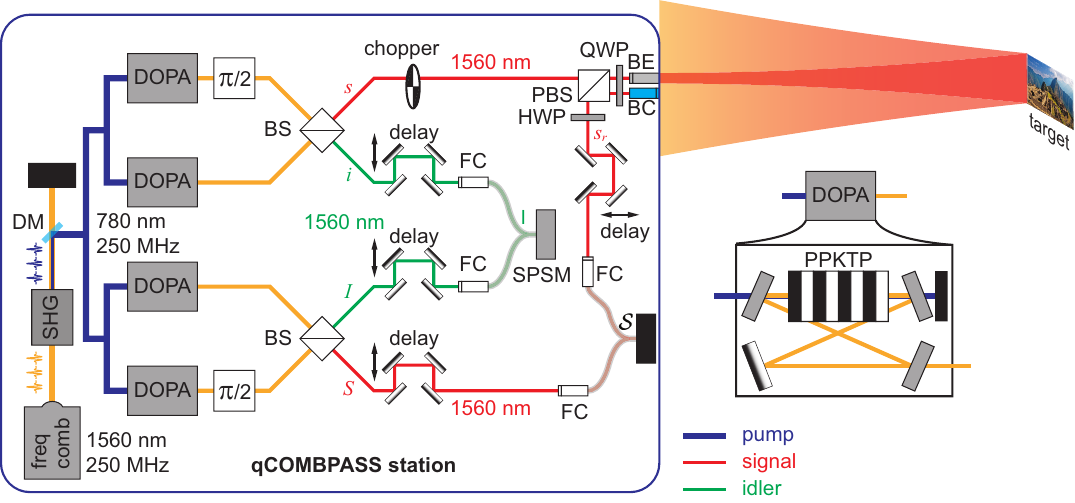}
\caption{{\bf Schematics of non-perturbative two-mode squeezed vacuum qCOMBPASS experimental setup.}
A classical frequency comb pumps four PPKTP degenerate optical parametric amplifiers (DOPAs) in a bow-tie resonator configuration operated below threshold. Target is in remote range (not to scale). Probe signal pulses of forward TMSVc
undergo diffraction, loss, and noise. 
BC (beam collector), 
BE (beam expander), 
BS (broad-band 50-50 beam splitter),
DM (dichroic mirror), 
FC (fiber coupler),
HPW (half wave plate),
PBS (polarizing beam splitter), 
PPKPT crystal (potassium titanyl phosphate),
QWP (quarter wave plate),
SHG (second harmonic generation), 
SPSM (single photon spectrometer).
}
\label{Fig6}
\end{figure}
%%%%%%%%%%%%%%%

We use the qCOMBPASS equation Eq.(\ref{qCOMBPASS}) to estimate photo-counts and interference visibility. 
Consider as an example a classical frequency comb with center wavelength $\lambda_c=1560$ nm, 
repetition frequency $\omega_{rep}/2\pi = 250$ MHz, carrier envelope offset stabilized at zero,
linear polarization, and $>200$ mW average output power. This output power is more than enough to pump
the four DOPAs at high squeezing levels 
(In previously reported work \cite{Vahlbruch2016} only 16 mW of second harmonic pump power was needed to reach 
15 dB squeezing in 1060 nm single-mode squeezed cw-light with a  doubly-resonant 
PPKTP OPA). The four DOPAs are optimized for  input/output central wavelengths (778,1560) nm and the bow-tie resonators have free spectral range 
$\omega_{FSR}/2\pi = 125$ MHz.
We assume  squeezed combs with
15 dB squeezing are produced with 125 MHz inter-teeth separation and Gaussian spectral weights with $\eta_{m=0} \approx 0.2$.
The rotational frequency of the chopper is $\sim 10$ kHz so that trains of $p \sim 10^4$ consecutive signal pulses can be blocked/unblocked. We consider a target at a distance $d=100$ km from the qCOMBPASS station
with  geometric cross section $\sigma_{geom} = 5\times 5 \; {\rm m}^2$ and
reflectance $r^2=0.5$ for all comb frequencies. 
We take into account beam diffraction by assuming the telecom signal probe pulse at the exit of the qCOMBPASS station has a Gaussian profile with a beam waist of 
$w_0=30$ cm and 
diffraction-limited beam divergence $\theta=1 \mu$rad, so the free-space diffraction  Rayleigh range is $z_0 \approx 180$ km
and the illumination spot size at target position is $w(d) \approx 34$ cm, implying coherent full illumination of the target.  The  atmospheric extinction is $l_m \approx (0.8)^2$  at telecom wavelengths 
\cite{Weichel}, background thermal noise is negligible at ambient temperature for visible/near-IR light, and beam spreading due to turbulence is 
assumed to be small compared with free-space diffraction \cite{Goodman1985}.  
We consider perfect spatiotemporal overlaps $O_{{\cal S}_m}=O_{{\cal I}_m}=1$ at the Y-junctions. 
The reflected signal beam collection efficiency is $\mu_{coll} \approx 0.65$ (ratio of beam collector area to
reflected spot size) and the
detection efficiency is $\mu_d=0.9$ (achievable with superconducting nanowire single-photon detectors (SNSPD) that also have $\le 1$ Hz dark counts at 1560 nm). 
For these parameters, we estimate a qCOMBPASS maximum photo-count rate of
$R_m \sim10^7$ idler photons/sec and a visibility $V_m \approx 40 \%$. 

%%%%%%%%%%%%
\vspace{0.2cm}
\section{Discussion}

qCOMBPASS breaks new ground for quantum remote sensing 
and embedded ranging applications.
Long intra- and inter-coherence times of quantum frequency combs, together with path quantum indistinguishability, enables quantum sensing and imaging of targets of interest at remote locations, without quantum memories and with robustness against photon loss, noise, and decoherence. As the scheme is basically a comb-based nonlinear quantum interferometer, the detected signal is governed by the interference between two electromagnetic quantum fields and depends linearly on the reflection coefficient of the target. This enables qCOMBPASS detection of very low-reflectivity distant objects with a square-root quantum advantage over incoherent LIDAR. The signal probe mode belonging to the two-mode squeezed comb is low-profile, involving only a few photons, which is conducive to applications where the goal is to detect without being detected. Detrimental effects of electromagnetic noise and thermal background can be substantially reduced by filtering out any photon trying to enter the qCOMBPASS station at wavelengths different from the comb frequencies of the signal pulse comb that flew to target. Background thermal noise has a negligible effect on qCOMBPASS performance for probe signal combs in the optical/infrared frequency range.  
A complete analysis of the effects of environmental noise and decoherence on the qCOMBPASS protocol will be the subject of a future study. 
As in standard ZWM schemes, qCOMBPASS offers some intrinsic level of security against spoofing and jamming attacks because it is based on nonlinear generation of entangled quantum frequency combs, and the corresponding parametric down-conversion or four-wave mixing 
phase matching conditions are not available to outside observers. Finally, qCOMBPASS has the potential to strongly impact remote quantum sensing, ranging, and imaging applications, including anti-stealth/anti-cloak technologies and low detection probability surveillance and hyperspectral imaging from space. The key idea behind qCOMBPASS should also be of interest for quantum communications, including secure time-transfer for clock synchronization in Earth-satellite links, distributed entanglement in multi-mode quantum networks, and VLB nonlinear interferometry  in space for quantum GPS. The experimental demonstration of qCOMBPASS is under way in our team. 

%%%%%

\vspace{0.2cm}
\noindent
{\bf Author Information}

\noindent
{\bf Notes} \\
The authors declare no competing interests. 

%%%%%

\vspace{0.2cm}
\noindent
{\bf Contributions.}
DD conceived qCOMBPASS and developed the theory.
TV contributed to the theoretical analysis of the path identity operation in the non-perturbative case.
YC conducted the experimental demonstration of ZWM with a cw laser.
AA made feasibility studies of experimental schemes.
HTC performed experiment-theory analysis.
PM did the full study of the qCOMBPASS equation and its distance scaling.
AA, HTC, and DD designed the experimental implementation.
All authors contributed to the writing of the manuscript.
%%%%%%

\vspace{0.2cm}
\noindent
{\bf Acknowledgments.}
We are grateful to J. Cox for insightful discussions on qCOMBPASS. We also acknowledge discussions with N. Dallmann, M. Everhart-Erickson, L.P. Garcia Pintos Barcia,
W. Kort-Kamp, A. L\'{o}pez Rubio, M. Lucero, R. Menzel, R. Newell,  A. Touil, M. Wallace, and W. Zurek. 
Special thanks go to Profs. M. Cho and T.Y. Yoon at Korea University for crucial insights into frequency comb
single-photon interferometry, and to Prof. M. Endo at University of Tokyo for discussions on photon squeezing.  
This work was supported by Los Alamos National Laboratory LDRD program. 
AA and HTC acknowledge partial support from the Center for Integrated Nanotechnologies.

%%%%%

\vspace{0.2cm}
\noindent
{\bf Data availability.}
Data and calculations are available upon request.

%%%%

%%%%%%

\vspace{0.2cm}
\noindent
{\bf References.}

%%%
\vspace{0.2cm}

\section
{\bf Appendix} 

We derive Eq. (\ref{qCOMBPASS}) of the main paper for the photo-counts at the idler detector.
After path identity, the relevant part of the qCOMBPASS state for the photo-count calculation  is 
Eq. (\ref{transceiver}) of the main paper, namely,
$\rho_{{\rm qCOMBPASS}} \cong \rho_{{\cal S}} \otimes \rho_{{\cal I}} \otimes \rho_{{\cal S} {\cal I}}
\otimes \rho_{E} \otimes \rho_{T} \otimes \rho_B \otimes \rho_{C}$, where
\begin{eqnarray}
\rho_{{\cal S}} &\cong&
\mathbb{U}_{B} \mathbb{U}_{T}\mathbb{U}_{E} 
\frac{1}{{\cal N}_s}
{\rm Tr}_{i} 
\left[
\Pi_{i} \mathbb{S}^{}_{{\cal S},i}\;
|0 \rangle \langle 0| \; 
\mathbb{S}^{\dagger}_{{\cal S},i}
\Pi_{i}
\right]  \mathbb{U}^{\dagger}_{B} \mathbb{U}^{\dagger}_{T} \mathbb{U}^{\dagger}_{E}
\nonumber \\
\rho_{{\cal I}} &\cong&
\mathbb{U}_{chop} \; \mathbb{S}^{}_{s,{\cal I}}\;
|0 \rangle \langle 0| \;
\mathbb{S}^{\dagger}_{s,{\cal I}}
\; \mathbb{U}^{\dagger}_{chop}
\nonumber \\
\rho_{{\cal S},{\cal I}} &\cong&
\mathbb{S}^{}_{{\cal S},{\cal I}}\;
|0 \rangle \langle 0| \;
\mathbb{S}^{\dagger}_{{\cal S},{\cal I}} 
\end{eqnarray}
We express the squeezing operators in terms of their expansion in the dual-Fock basis and perform the measurement and
unitary operations to get
\begin{eqnarray}
\rho_{{\cal S}} &=& 
\frac{1}{\cosh^2[\eta \tilde{g}(d) ]} \sum_{q,q'=0}^{\infty}e^{i (q-q')(\varphi_{s_r}+\phi_r+ \phi_g+\phi_{\xi})}
\{ \tanh[\eta \tilde{g}(d) ] \} ^{q+q'}  |q\rangle_{\cal S} \langle q'|_{\cal S} 
\nonumber \\
\rho_{{\cal I}} &=&\frac{1}{\cosh^2(\eta g)} \sum_{p,p'=0}^{\infty}e^{i (p-p')(\varphi_{i}+\phi_g)}
[\tanh(\eta g)]^{p+p'}  |p\rangle_{\cal I} \langle p'|_{\cal I} 
 \nonumber  \\
\rho_{{\cal S} {\cal I}} &=&\frac{1}{\cosh^2(\eta g )} \sum_{n,n'=0}^{\infty}e^{i (n-n')(\varphi_{S}+\varphi_{I}+\phi_g)}
[\tanh(\eta g)]^{n+n'}  |n\rangle_{{\cal S}} \langle n'|_{{\cal S}} \otimes |n\rangle_{{\cal I}} \langle n'|_{{\cal I}} 
\end{eqnarray}
where we defined the renormalized squeezing amplitude 
$\tilde{g}(d)=g \mu_{coll} r \ell z_0/d<g$.
The first line is for a signal comb in a 
single mode , the second line is for an idler comb in a single mode, 
and the third line is for a signal-idler bicomb in a two-mode squeezed state (TMSVc). Hence, after path identity in the corresponding Y-junctions,
\begin{eqnarray}
\rho_{{\rm qCOMBPASS}} &\cong& 
\frac{1}{\cosh^4(g \eta) \cosh^2(\tilde{g}\eta)}
\sum_{n,n'=0}^{\infty} \sum_{p,p'=0}^{\infty} \sum_{q,q'=0}^{\infty}  \nonumber \\
& \times &
e^{i [(n-n')(\varphi_{S}+\varphi_{I}+\phi_g)+ (p-p')(\varphi_{i}+\phi_g)+ (q-q')(\varphi_{s_r}+\phi_r+ \phi_g+\phi_{\xi})]}
\nonumber\\
&\times & [\tanh(g \eta)]^{n+p+n'+p'} [\tanh(\tilde{g} \eta)]^{q+q'}  
\nonumber \\
& \times &
|n+q\rangle_{{\cal S}} \langle n'+q'|_{{\cal S}} \otimes |n+p\rangle_{{\cal I}} \langle n'+p'|_{{\cal I}} 
\label{rhocombpass}
\end{eqnarray}
To ensure convergence of the sums, one must impose $\tanh(g \eta)<1$ and $\tanh(\tilde{g} \eta)<1$, and care must be exercised when taking the strong non-perturbative limit in denominators. 
The photo-count at the idler photons is $N_{\cal I} =  {\rm Tr} [ a^{\dagger}_{\cal I}  a_{\cal I}  \rho_{{\rm qCOMBPASS}}]$. Only the diagonal terms
$n+q=n'+q'$ and $n+p=n'+p'$ contribute:
\begin{eqnarray}
N_{\cal I}= (1-x^2)^2 (1-y^2)
\sum_{n,p=0}^{\infty} (n+p) x^{2(n+p)}  \sum_{q,q'=0}^{\infty} y^{q+q'} e^{i (q-q') \Phi} 
\end{eqnarray}
where $x=\tanh( g\eta)$,  $y=\tanh( \tilde{g}\eta)$, and
$\Phi=\varphi_{s_r}+\varphi_{i} - \varphi_{S}-\varphi_{I}+\phi_r+ \phi_g+\phi_{\xi}$.
We separate the terms $q=q'$ (baseline) and $q \neq q'$ (interference):
$N_{\cal I}=N^{(baseline)}_{{\cal I}} + N^{(interf)}_{{\cal I}}$, and find
\begin{eqnarray}
N^{(baseline)}_{{\cal I}} &=&
(1-x^2)^2 (1-y^2)
\sum_{n=0}^{\infty}  x^{2n} \sum_{p=0}^{\infty}  (n+p) x^{2p} 
\sum_{q=0}^{\infty} y^{2q} 
\nonumber \\
&=&
(1-x^2)^2 (1-y^2) 
\sum_{n=0}^{\infty}  x^{2n} \left[ \frac{n}{1-x^2} + \frac{x^2}{(1-x^2)^2} \right]
\frac{1}{1-y^2}
\nonumber \\
&=&
(1-x^2)^2  \left[ \frac{x^2}{(1-x^2)^3} + \frac{x^2}{(1-x^2)^3} \right]
\nonumber \\
&=&
\frac{2x^2}{1-x^2}
\end{eqnarray}
For the interference part, $q \neq q'$,
\begin{eqnarray}
N^{(interf)}_{{\cal I}} &=&
(1-x^2)^2 (1-y^2) \sum_{n=0}^{\infty} x^{2n} \sum_{p=0}^{\infty} (n+p) x^{2p}  \sum_{q=1}^{\infty}  (y e^{i \Phi})^q
\sum_{q'=0}^{q-1} (y e^{-i \Phi})^{q'} + c.c.
\nonumber \\
&=&
\frac{2x^2 (1-y^2) }{1-x^2}  \sum_{q=1}^{\infty}  (y e^{i \Phi})^q
\frac{1-(y e^{-i \Phi})^{q}}{1-y e^{-i \Phi}} + c.c.
\nonumber \\
&=&
\frac{2x^2 (1-y^2)}{1-x^2} \frac{1}{1-y e^{-i \Phi}}
\sum_{q=1}^{\infty} [ (y e^{i \Phi})^q -y^{2q} ] + c.c.
\nonumber \\
&=&
\frac{2x^2 (1-y^2) }{(1-x^2) (1+y^2-2y \cos\Phi)} (1-y e^{i \Phi})
\left[ \frac{y e^{i \Phi}}{1-y e^{i \Phi}} - \frac{y^2}{1-y^2} \right] + c.c.
\nonumber \\
&=&
\frac{2x^2 (1-y^2) }{(1-x^2) (1+y^2-2y \cos\Phi)} 
\left[y e^{i \Phi} - \frac{y^2 }{1-y^2} + \frac{y^2 e^{i\Phi}}{1-y^2} \right] + c.c.
\nonumber \\
&=&
\frac{4x^2  }{(1-x^2) (1+y^2-2y \cos\Phi)} (y \cos \Phi - y^2)
\nonumber \\
&=&
\frac{2 x^2}{(1-x^2)} \; \frac{2y (\cos \Phi-y)}{1+y^2-2y \cos \Phi}
\end{eqnarray}
Hence, 
\begin{eqnarray}
N_{{\cal I}} &=&
\frac{2x^2}{1-x^2} 
\left[
1+
\frac{2y (\cos \Phi- y)}{1+y^2-2y \cos\Phi}
\right]
\end{eqnarray}
Finally, we introduce possible time-delays between pulses within the qCOMBPASS station (controlled by delay lines, see Fig. 3 of the main text)
\begin{eqnarray}
O_{{\cal I}} &=&  \int f^*_{i}({\bf x},t) f_{I}({\bf x},t-\tau_{i,I}) d^3x \nonumber \\
O_{{\cal S}} &=& \int f^*_{s_r}({\bf x},t-\tau_{i,s_r}) f_{S}({\bf x},t-\tau_{i,s_r}-\tau_{s_r,S}) d^3x
\end{eqnarray}
where  $t$ is a detection time and $\tau_{i,I}, \tau_{i,s_r},\tau_{s_r,S}$ are time-delays between indicated pulses. The qCOMBPASS equation is then
\begin{eqnarray}
N_{{\cal I}}&=&
2 \mu_d \sinh^2(g \eta) 
\left\{
1+
\frac{2 O_{{\cal S}} O_{{\cal I}}  \tanh(\tilde{g} \eta) [\cos \Phi- \tanh(\tilde{g} \eta)]}
{1+\tanh^2(\tilde{g} \eta)-2 \tanh(\tilde{g} \eta) \cos\Phi}
\right\}
\end{eqnarray}
where $\mu_d$ is the detector's efficiency. 
The visibility of the interference pattern is
\begin{eqnarray}
V_{{\cal I}} \!=\!
\frac{N^{(max)}_{{\cal I}}\!-\!N^{(min)}_{{\cal I}}}{N^{(max)}_{{\cal I}}\!+\!N^{(min)}_{{\cal I}}}
=  \frac{2 O_{{\cal S}} O_{{\cal I}}  y}{1+ O_{{\cal S}} O_{{\cal I}}  y^2}=
\frac{2 O_{{\cal S}} O_{{\cal I}}   \tanh(\tilde{g} \eta) }{ 1+ O_{{\cal S}} O_{{\cal I}}  \tanh^2(\tilde{g} \eta)}
\end{eqnarray}
Assuming a Gaussian probe pulse at the exit of the qCOMBPASS station with carrier wavelength $\lambda_s$ and beam waist $w_0$,
the Rayleigh range for free-space propagation is $z_0=\pi w_0^2/\lambda_{m=0}$ and
the transverse extent of the field is $w(z)=w_0 (1+ z^2/z_0^2)^{1/2}$. 
Each comb tooth suffers slightly different diffraction according to its wavelength,
$\lambda_m= 2\pi c/\omega_{s,m}$.
The illumination spot at the target position is $w(d)$. On the return, 
we take as new initial beam waist $w'_0 \approx w(d)$, the new Rayleigh range is
$z'_0=\pi w_(d)^2/\lambda_{m=0}$, and the reflected illumination
spot at the qCOMBPASS base position is 
$w'(0)=w'_0 (1+ d^2/z_0^{'2})^{1/2}$.
Assuming full illumination of the target (target geometric cross-section larger than $w(d)$)
the renormalized squeezing parameter is approximated as
$\tilde{g}=g \mu_{coll} r \ell z_0/d$. 
For parameters
$g=1.7$ (15 dB squeezing), 
$\eta_{m=0}=0.2$, 
$\lambda_{m=0}=1560$ nm,
$r^2=0.5$ (Si target), 
$\ell \approx 0.8\times 0.8=0.64$ (roundtrip atmospheric loss),
$w_0=30$ cm (achieved via a telescopic beam expander), and
$d=100$ km, 
we get
$z_0 \approx 180 \; {\rm km}$,
$w(d) \approx  34\; {\rm cm}$, 
$z'_0 \approx 230 \; {\rm km}$,
and $w'(0) \approx  37\; {\rm cm}$.
The collection efficiency is $\mu_{coll}=[w_0/w'(0)]^2 \approx 0.65$. Hence
$\tilde{g} \eta_{m=0}\approx 0.2$.

\end{document}